\documentclass[12pt]{iopartF}

 \expandafter\let\csname equation*\endcsname\relax
 \expandafter\let\csname endequation*\endcsname\relax

\usepackage{amscd,amssymb,amsmath,amsthm}

\usepackage{epsfig}
\usepackage{graphicx,xspace}
\usepackage{subfig }
\usepackage{pstricks}
\usepackage[left=3cm,top=2.9cm,right=3cm,bottom=2.9cm]{geometry}
\usepackage{color}
\usepackage{comment}
\usepackage{amsmath}
\definecolor{labelkey}{cmyk}{.4,.2,0,0}
\usepackage{amsfonts}
\usepackage{amsthm}
\usepackage{commath}
\usepackage{cases}
\usepackage{dsfont}
\usepackage{cite}
\theoremstyle{remark}

\renewcommand*\contentsname{}

\begin{document}

\newcommand \be  {\begin{equation}}
\newcommand \ee  {\end{equation}}
\newcommand \bea {\begin{eqnarray} }
\newcommand \eea {\end{eqnarray}}

\title{Simple derivation of the $(- \lambda H)^{5/2}$ tail for the
1D KPZ equation.}

%\title{Simple derivation of the long-time large deviations for the height of the 1D KPZ equation with Brownian and droplet initial conditions for $\lambda H<0$.}
%

\vskip 0.2cm

\author{Alexandre Krajenbrink and Pierre Le Doussal }

\address{CNRS-Laboratoire de Physique Th\'eorique de l'Ecole Normale Sup\'erieure\\
24 rue Lhomond, 75231 Paris
Cedex-France}

\date{Received:  / Accepted:  / Published }
% The correct dates will be entered by the editor

\begin{abstract}
We study the long-time regime of the Kardar-Parisi-Zhang (KPZ) equation in $1+1$ dimensions for the Brownian and droplet initial conditions and present a simple derivation of the tail of the large deviations of the height on the negative side $\lambda H<0$. We show that for both initial conditions, the cumulative distribution functions take a large deviations form, with a tail for $- \tilde s \gg 1$ given by $-\log \mathbb{P}\left(\frac{H}{t}<\tilde{s}\right)=t^2 \frac{4 }{15 \pi} (-\tilde{s})^{5/2} $. This exact expression was already observed at small time for both initial conditions suggesting that these large deviations remain valid at all times. We present two methods to derive the result (i) long time estimate using a Fredholm determinant formula and (ii) the evaluation of the cumulants of a determinantal point process where the successive cumulants appear to give the successive orders of the large deviation rate function in the large $\tilde s$ expansion.
An interpretation in terms of large deviations for trapped fermions at low temperature is
also given. In addition, we perform a similar calculation for the KPZ equation in a half-space 
with a droplet initial condition, and show that the same tail as above arises, with the prefactor $\frac{4}{15\pi}$ replaced by $\frac{2}{15\pi}$. Finally, the arguments can be extended to show that this tail holds for all times. This is consistent with the fact that the same tail was obtained previously in the short time limit for the full-space problem. 
%We also extend this result to the half-space by calculating the full large deviation function $\Phi(H)$ in the short time limit.
\end{abstract}
\maketitle
\tableofcontents
%\newpage
\section{Introduction and main results}
\label{introduction}
Many works have been devoted to studying 
the 1D continuum KPZ equation \cite{KPZ, directedpoly,  HairerSolvingKPZ,  reviewCorwin, HH-TakeuchiReview} 
which describes the stochastic growth of an interface of height $h(t,x)$ at point $x$ and time $t$ as
\be
\label{eq:KPZ}
\partial_t h = \nu \, \partial_x^2 h + \frac{\lambda_0}{2}\, (\partial_x h)^2 + \sqrt{D} \, \xi(t,x) \;,
\ee
starting from a given initial condition $h(t=0,x)$.
Here
%$\nu > 0$ denotes the strength of the diffusive relaxation, $\lambda_0 > 0$ is the coefficient of the 
%non-linearity and 
$\xi(x,t)$ is a centered Gaussian white noise with 
$\mathbb{E} \left[\xi(t,x) \xi(t',x')\right]= \delta(x-x')\delta(t-t')$, and we
use from now on units of space, time and heights such
that $\lambda_0=D=2$ and $\nu=1$ \cite{footnote100,footnote1}. \\

Recently, a large research effort has been devoted to the \textit{large deviations} away from the typical behavior. Unlike diffusive interacting particle systems for which powerful methods \cite{Derrida_largedeviation, Derrida_MFT} were developed, systems in the KPZ class require new theoretical efforts. Recently, the large deviations behavior of the KPZ equation has been investigated at short times  \cite{Baruch,le2016exact,MeersonParabola,janas2016dynamical, meerson2017randomic, krajenbrink2017exact} (see also \cite{CLR10,flatshorttime}) and, to a lesser extend, in the long time limit \cite{LargeDevUs, sasorov2017large}. One outstanding open question is whether the tails of the large deviations remain valid at all times and how they depend on the initial condition. \\

Our interest here lies in the left tail ($H<0$) of the distribution of the shifted solution, 
$H(t)=h(t,x=0)+\frac{t}{12}$,  and particularly in its large negative $H$ behavior.
In previous works it was shown for flat \cite{Baruch}, for droplet \cite{le2016exact, MeersonParabola} and Brownian \cite{janas2016dynamical,krajenbrink2017exact,meerson2017randomic} initial conditions (IC), that {\it in the short time limit}, $t\ll 1$ and $H$ fixed, the distribution of $H$ takes the large deviation form  $P(H,t)\sim e^{-\frac{\phi(H)}{\sqrt{t}}}$, with the leading asymptotics
\be
\label{largedev111}
\phi(H) \simeq_{H \to -\infty} \frac{4}{15\pi}(-H)^{5/2}
\ee
identical for all these IC. There also exist results {\it for the large time limit} $t\gg 1$, where in the regime $H/t$ fixed, the distribution of $H$ takes the form 
\be
P(H,t)\sim e^{-t^2 \Phi_-(\frac{H}{t})} \label{left} 
\ee
Following the work of Ref. \cite{LargeDevUs} and using a rather involved analysis of a formidable-looking non-local Painlev\'e (NLP) equation (originally derived in Ref. \cite{ACQ11}), an explicit formula for $\Phi_-(\frac{H}{t})$ for droplet IC was obtained in Ref. \cite{sasorov2017large}
\begin{equation}
\label{eq_Meerson}
\Phi_-(\tilde{s})=\frac{4}{15\pi^6}(1-\pi^2 \tilde{s})^{5/2}-\frac{4}{15\pi^6}+\frac{2}{3\pi^4}\tilde{s}-\frac{1}{2\pi^2}\tilde{s}^2
\end{equation}
We stress that this remarkable result could be derived only in the case of the droplet IC, where
the NLP equation is available.

In this paper, we are able to treat several initial conditions - the droplet and the Brownian IC.
Our goal is more modest, as we obtain only the large negative
$\tilde s$ asymptotics of the rate function $\Phi_-(\tilde s)$, 
however our derivations are much simpler and, we hope, more versatile. In addition we obtain a systematic method to calculate $\Phi_-(\tilde s)$,
in a large $\tilde s$ expansion, yet to be fully exploited.

We use two different and complementary methods (i) the large time estimate of the Fredholm determinant associated to the solution of the KPZ equation and (ii) the evaluation of the cumulants of a determinantal point process associated to this Fredholm determinant. We find, for both IC, that the cumulative distribution exhibits the large deviation tail 
\begin{equation}
\label{large_dev}
-\log \mathbb{P}(H<t\tilde{s})=t^2\frac{4}{15\pi}(-\tilde{s})^{5/2}
\end{equation}
 for large negative $\tilde{s}$, consistent with Eq. \eqref{eq_Meerson}. This result is obtained through a very simple derivation which we think is worth presenting in detail here. We conjecture and verify up to third order on the droplet IC, that the cumulant expansion gives the large negative $\tilde{s}$ series expansion of the large deviation rate function $\Phi_-$ in Eq. \eqref{eq_Meerson}, where the $n$-th cumulant gives the $n$-th order of the expansion. In addition, we perform a similar calculation for the KPZ equation in a half-space and we show that the same tail as in Eq. \eqref{largedev111} arises with the prefactor $\frac{4}{15\pi}$ replaced by $\frac{2}{15\pi}$. Finally, the arguments can be extended to show that this tail holds for all times, which is consistent with the fact that the same tail was obtained previously in the short time limit. The cumulant expansion method
allows to obtain exact bounds for the left tail, which are saturated at short time and
allow to bound the large deviation rate function at large time.
 
Note that a simultaneous and independent mathematical work also studies the left tail of the KPZ equation for droplet initial conditions \cite{corwin2018lower}. A summary combining their results as well as the present results will appear shortly \cite{AllofUs}. The method developed here is also applied to obtain exact results for short time probability distribution for a variety of cases in \cite{UsHalfSpace}. Finally, a very recent numerical work, which probes the tails of the KPZ equation with droplet initial condition using a lattice directed polymer representation, shows good agreement with our predictions \cite{hartmann2018high}. 
\\

The outline of the paper is as follows. We start in Section \ref{recall} and \ref{choice_scaling} by recalling the exact starting formulae and the proper choice of scaling in the large time large deviation regime. The two independent derivations are given in two different sections, respectively \ref{method1} and \ref{method2}. In Section \ref{method1}, we start from the exact Fredholm determinant formula for the moment generating function of the solution of the KPZ equation and derive its large time trace expansion to provide the long time large deviation rate function. Remarkably the very same method was used to derive the \textit{short time} large deviations. A simple identification of the short time and large time limit of the kernels of the Fredholm determinants explains why the same method surprisingly work for both regimes. In Section \ref{method2}, we provide a more powerful method based on the evaluation of the cumulants of a determinantal point process where the dominant term, which gives Eq. \eqref{large_dev}, only involves the density of states of the related process. In the case of the Airy process associated to the droplet IC, Eq. \eqref{large_dev} can be obtained by considering the edge of Wigner's semi-circle. Following the derivation of the second method, we discuss its range of validity. In Section \ref{alltimes}, we discuss its application to earlier times to show that the tail holds for all times. In Section \ref{half_space_extension}, we argue that it can be applied to a Pfaffian point process : the KPZ half-space problem. In Section 8, we obtain exact bounds from (i) the first cumulant and (ii) the
conditional density, whose consequences are discussed. 
 In Section \ref{fermion_large_dev}, we interpret our results in terms of trapped fermions and we finally present in the Appendices more details about the calculations of the main text and a conjecture about a relation between the cumulant expansion and the Baker-Campbell-Hausdorff formula.
\section{Recall of exact results}
\label{recall}
We recall the exact results valid at all times for the two initial conditions studied here, which provide a starting formula for the analysis in Section \ref{method1}.
\begin{enumerate}
\item \textit{Droplet initial condition } \\
The droplet initial condition is $h_{\mathrm{drop}}(t=0,x)=-\frac{\abs{x}}{\delta}-\ln(2\delta)$ with $\delta \ll 1$ and the moment generating function of $e^H$ is then given by, see \cite{SS10, ACQ11, DOT10, CLR10}, 
\begin{equation}
\label{fredholm1}
\mathbb{E}_{\, \mathrm{KPZ}}\left[\exp\left(-e^{H(t)-st^{1/3}}\right) \right]=Q_t(s)=\mathrm{Det}\left[ I-\bar K_{t,s}\right]
\end{equation}
where the average is taken with respect to the KPZ white noise. $Q_t(s)$ is a Fredholm determinant associated to the kernel 
\begin{equation}
\bar K_{t,s}(u,u') = K_{\rm Ai}(u,u') \sigma_{t,s}(u')
\end{equation}
defined in terms of the Airy kernel and the weight function
\begin{equation}
\begin{split}
&K_{\mathrm{Ai}}(u,u')=\int_{0}^{+\infty} \! \mathrm{d}r \; \mathrm{Ai}(r+u) \mathrm{Ai}(r+u') \\
&\sigma_{t,s}(u)=\sigma(t^{1/3}	(u-s)) \quad , \quad \sigma(v)=\frac{1}{1+e^{-v}}
\end{split}
\end{equation}
\item \textit{Brownian initial condition} \\
The Brownian initial condition with drift is $h_{\mathrm{brownian}}(t=0,x) = B(x) - w |x|$, where $B$ is a double-sided Brownian motion. For this initial condition one needs to introduce a real random variable $\chi$ independent of $H$, with a probability density 
$p(\chi) d\chi =  e^{-2 w \chi - e^{-\chi}} d\chi/ \Gamma(2 w)$ so that the moment generating function is given by, see Refs. \cite{SasamotoStationary,SasamotoStationary2,BCFV},
\begin{equation}
\mathbb{E}_{\, \mathrm{KPZ},\, B, \, \chi}\left[\exp\left(-e^{H(t)+\chi-st^{1/3}}\right) \right]=Q^\Gamma_t(s)=\mathrm{Det}\left[ I-\bar K^{\Gamma}_{t,s}\right]
\end{equation}
where the average is taken over the KPZ noise, the random initial condition and the random variable $\chi$. $Q^\Gamma_t(s)$ is a Fredholm determinant associated to
the kernel
\be \label{Kt} 
\bar  K^\Gamma_{t,s}(v,v') = K_{\rm Ai, \Gamma}(v,v')\sigma_{t,s}(v')  \ee \vspace{-0.2cm}
defined in terms of the weight function and the deformed Airy kernel
\be \label{deformedK} \begin{split} 
  K_{\rm Ai, \Gamma}(v,v') = &\int_{0}^{+\infty} dr \, \mathrm{Ai}_\Gamma^\Gamma(r+v,t^{-\frac{1}{3}},w,w) \mathrm{Ai}_\Gamma^\Gamma(r+v',t^{-\frac{1}{3}},w,w) 
\end{split} \ee 
 itself defined from the deformed Airy function 
\be  \label{aigamma} 
\mathrm{Ai}_\Gamma^\Gamma(a,b,c,d) := \frac{1}{2 \pi}\int\limits_{-\infty+i \epsilon}^{+\infty+i \epsilon}\mathrm{d}\eta \; \mathrm{exp}   (i \frac{\eta^3}{3}+ i a\eta)\frac{\Gamma(i b\eta+d)}{\Gamma(-i b\eta+c)}  
\ee
where $\epsilon \in [0, \mathrm{Re}(d/b)[$ and $\Gamma$ is the  Gamma function.
\end{enumerate}
\section{Choice of scaling at large time and large deviation function}
\label{choice_scaling}
It is known that, at large time, the typical fluctuations of the height are of order one third,  $H(t)\simeq \Upsilon t^{1/3}$ with a random variable $\Upsilon$ which depends on some broad features of the initial condition. In the large deviation regime the scaling of fluctuations is different and is actually linear in time as observed in Refs. \cite{LargeDevUs, sasorov2017large, meerson2017randomic}. Since the moment generating function has argument $\exp\left(-e^{H(t)-st^{1/3}}\right) $, a convenient scaling is \vspace{-0.15cm}
\be
H(t) =\tilde{H} t \quad , \quad s=\tilde{s}t^{2/3}
\ee 
  with $\tilde{H}$  and $\tilde{s}$ of order one. The reason for this is that the argument of the average then becomes $\exp\left(-e^{t(\tilde H-\tilde s)}\right)$ where $\tilde{H}$ and $\tilde{s}$ are of the same order. The weight function having the form $\sigma(t^{1/3}	(u-s))$, using the already rescaled $s$, we see that the same rescaling is required for $u$, therefore we define $u=\tilde{u}t^{2/3}$, with $\tilde{u}$ of order one.\\
~\\
Regarding the random variable $\chi$, we also need to rescale it as $\chi=\tilde \chi t$ and the probability measure then becomes
\begin{equation}
e^{-2 w \chi - e^{-\chi}} d\chi/ \Gamma(2 w)=t e^{-2 wt \tilde  \chi - e^{-t\tilde \chi}}  d\tilde \chi/ \Gamma(2 w)
\end{equation}
The factor $e^{- e^{-t\tilde \chi}} $ becomes at large time $\mathds{1}_{\tilde{\chi}\geq 0}$, leading, after renormalization of the measure, to an exponential distribution $P(\tilde{\chi})\mathrm{d}\tilde{\chi}=2wt  e^{-2 wt \tilde  \chi }\mathds{1}_{\tilde{\chi}\geq 0}\mathrm{d}\tilde{\chi}$. Besides, at fixed $w$, this distribution is a nascent delta function, leading to $\lim_{t\to \infty} P(\tilde{\chi})=\delta(\tilde{\chi})$. The conclusion of this is that we can discard $\chi$ at large time and set it to 0.\\
~\\
This choice of scaling is convenient to introduce the large deviation function that we are interested in. Indeed, the generating function $\mathbb{E}_{\, \mathrm{KPZ}}\left[\exp(-e^{t(\tilde H-\tilde s)})\right]$ converges at large time to $\mathbb{P}(\tilde{H}<\tilde{s})$.
%\newpage
\section{Method A : long time estimate of Fredholm determinants}
\label{method1}
\subsection{Asymptotics of the (deformed) Airy kernel at large time }
We now claim that the large time asymptotics of the kernel for both initial conditions is identical, giving an equivalence between these initial conditions in the large time estimates and prove it in the Appendix in Section \ref{details_calculation_sine}. The asymptotics of both the deformed Airy kernel and the Airy kernel are given by, for $v<0$
\begin{equation}
 \begin{split} \label{KtAi} 
  K(vt^{2/3},(v+\frac{\kappa}{t})t^{2/3})
  &\underset{t\gg 1}{\simeq} \frac{ t^{1/3}}{\pi } \frac{\sin (\sqrt{-v} \kappa)}{\kappa} \quad , \quad 
  K= K_{\rm Ai}, K_{\rm Ai, \Gamma}
\end{split} 
\end{equation}
i.e. in terms of the {\it sine-kernel.} We expect from the shared asymptotics that  the tail of the large deviation behavior that we are interested in will be identical for both initial conditions. It is therefore necessary to consider only one initial condition, and we choose the droplet one.
\subsection{Long time estimate of the Fredholm determinant related to the Airy kernel}
We will use in this Section the same method that was already used in Refs. \cite{le2016exact,krajenbrink2017exact} to derive the large deviations rate function for the solution of the KPZ equation at short time by means of evaluating the successive traces of the kernel of the Fredholm determinant. This suggests a common method for both short and long time, strengthening the idea of universal results valid in these two regimes.\\
\\
Using the identity $\log \mathrm{Det}=\mathrm{Tr}\log$, and expanding the logarithm into a series, the Fredholm determinant in Eq. \eqref{fredholm1} can be computed as 
\bea\label{Q_start_supp}
\log Q_t(s) = - \sum_{p=1}^\infty \frac{1}{p} {\rm Tr}\, \bar{K}^p_{t,s} 
\eea
where we recall the definition of the trace of a Kernel
\begin{equation}
\label{trace_kernel}
{\rm Tr} \; {\bar K}_{t,s}^p = \int_{-\infty}^\infty dv_1 \int_{-\infty}^\infty dv_2 \ldots \int_{-\infty}^\infty dv_p K_{\rm Ai}(v_1,v_2) \dots K_{\rm Ai}(v_p,v_1)  \sigma_{t,s}(v_1) \ldots  \sigma_{t,s}(v_p)
\end{equation}
The first approximation to use is the large time limit of the Airy kernel for $v<0$
given in \eqref{KtAi}.
%\begin{equation}
% \begin{split}
%  K_{\rm Ai}(vt^{2/3},(v+\frac{\kappa}{t})t^{2/3})
%  &\underset{t\gg 1}{\simeq} \frac{ t^{1/3}}{\pi } \frac{\sin (\sqrt{-v} \kappa)}{\kappa} 
%\end{split} 
%\end{equation}
For $v > 0$, the Airy kernel vanishes exponentially in thatlimit
and therefore only the region where all the $v$'s are negative needs to be considered
in \eqref{trace_kernel}. We define the relative coordinates $v_j=v_{j-1}+ \frac{\sqrt{|v_1|} 
\kappa_j}{t}$ and reduce the range of integration of the $v $'s due to the fact that the weight function vanishes exponentially for $v<\tilde{s}$.
\be
\begin{split}
&{\rm Tr} \bar{K}^p 
\simeq \frac{t}{\pi^p}   \int_{\tilde{s}}^0 dv_1 \sqrt{|v_1|} \int_{t\tilde{s}\sqrt{|v_1|}}^{-t\tilde{s}\sqrt{|v_1|}} d\kappa_1  \ldots \int_{t\tilde{s}\sqrt{|v_1|}}^{-t\tilde{s}\sqrt{|v_1|}} d\kappa_p\\
& \prod_{i=1}^p \sigma(t(v_1-\tilde s)+\sum_{j=2}^{i}\frac{\kappa_j}{\sqrt{|v_1|}})   \frac{\sin ( \kappa_1)}{\kappa_1} \frac{\sin (  \kappa_2)}{\kappa_2} \ldots \frac{\sin ( \kappa_p)}{\kappa_p}
\delta(\kappa_1+\kappa_2+\dots+\kappa_p) 
\end{split}
\ee
We now want to evaluate the leading order of the Fredholm determinant. We make the \textit{hypothesis} that we can neglect the terms $\sum_{j=2}^{i}\frac{\kappa_j}{\sqrt{|v_1|}}$ in the Fermi factors, and the limit in the $\kappa$ integrals can be extended to infinity at large time to give the dominant order of the Fredholm determinant. Under these hypothesis, the traces are simplified onto
\bea{\rm Tr} \bar{K}^p 
\simeq \frac{t}{\pi}   \int_{\tilde{s}}^0 dv_1 \sqrt{|v_1|}  \sigma(t(v_1-\tilde s))^p 
\eea
More details are given in the Appendix in Section \ref{appendix2} with arguments to obtain the next orders and corrections to the approximations made. This expression allows a simple summation over $p$
\vspace{-0.5cm }
\begin{equation}
\begin{split}
\log Q_t(s)&=-\frac{t}{\pi} \int_{\tilde{s}}^0 dv_1 \sqrt{|v_1|}  \sum_{p=1}^\infty \frac{ \sigma(t(v_1-\tilde s))^p}{p}\\
&\simeq -\frac{4t^2§}{15\pi}(-\tilde{s})^{5/2}
\end{split}
\end{equation} 
Hence we obtain our main result \eqref{large_dev}, valid for both IC. 
It coincides with the dominant order at large negative $\tilde s$ 
in the large deviations rate function \eqref{eq_Meerson} obtained in Ref. \cite{sasorov2017large}.
The conclusion of the present trace method for the Fredholm determinant is that evaluating all Fermi factors with the same argument and using the sine-kernel approximation for the Airy kernel yields the leading term $\log Q_t(s)=-\dfrac{4t^2}{15\pi}(-\dfrac{s}{t^{2/3}})^{5/2}$. This conclusion holds for both droplet and Brownian initial conditions as their kernel share the same asymptotics. Although we give some hints in the
Appendix, this method does not seem too easy to extend to obtain the next orders in the
large $-\tilde s$ expansion. We now turn to a second method, equally simple, and which appears more
convenient for that aim.
\\
%\newpage
\section{Method B : calculation of the cumulants of a determinantal point process and large time estimates}
\label{method2}
The goal of this section is to reproduce the calculation of the large deviations by computing the cumulants of a determinantal point process. 
\subsection{Introduction and aim of the method}
The method we present here is exact and starts from the observation that for both droplet 
and Brownian initial conditions, the kernel of the Fredholm determinant is written in the form $\bar K=K \sigma$ where $\sigma$ is the Fermi weight function, independent of the initial condition, whereas the kernel $K$ depends on the initial condition, $K_{\rm Ai}$ or $K_{\rm Ai, \Gamma}$ in our cases of interest. The method is valid for any $K$ (and can be extended to cases where $K$ depends on time).  We now suppose that the moment generating function of the solution of KPZ is associated to a Fredholm determinant with a kernel $\bar K=K\sigma$ so that we have the identity
\begin{equation} \label{KPZdet} 
\mathbb{E}_{\, \mathrm{KPZ}}\left[ \exp\left(  -e^{H(t)-st^{1/3}}\right) \right]=\mathrm{Det}\left[1-K \sigma\right].
\end{equation}
We recall the result of Refs. \cite{johansson2005random,johansson2015gaussian,borodinDeterminantalReview,borodin2016moments} that for a set of points $\lbrace a_i \rbrace_{i \in \mathbb{N}}$ following a determinantal point process with kernel $K$, we have in our system of unit
\begin{equation}
\label{deter_process}
\mathbb{E}_{K}\left[ \prod_{i=1}^{\infty} \frac{1}{1+e^{t^{1/3}(a_i-s)}}\right]=\mathbb{E}_{K}\left[ \prod_{i=1}^{\infty} (1-\sigma_{t,s}(a_i))\right]=\mathrm{Det}\left[1-K \sigma\right]
\end{equation}
The last identity being a standard property of generic determinantal point processes. We now define the quantity 
\be
\varphi_{t,s}(a)=-\log(1-\sigma_{t,s}(a))
\ee 
drop the subscript on $\varphi$ for simplicity and rewrite Eq. \eqref{deter_process} as
  \begin{equation}
 \begin{split}
\log \mathbb{E}_K \left[ \exp\left(-\sum_{i=1}^{\infty} \varphi(a_i)\right)\right]&=\log \mathrm{Det}(1-(1-e^{-\varphi})K)\\
&=\mathrm{Tr}\log (1-(1-e^{-\varphi})K)\\
&= -\sum_{p=1}^{+\infty} \frac{1}{p}\mathrm{Tr}[(1-e^{-\varphi})K]^p\\
\end{split}
 \end{equation}
Proceeding to the series expansion $1-e^{-\varphi}=\varphi-\dfrac{\varphi^2}{2}$ and going to the second order allows us to obtain the first two cumulants. Indeed, up to order two in $\varphi$, we have
 \begin{enumerate}
 \item \textit{p=1 }
 \begin{equation}
-\mathrm{Tr}[(1-e^{-\varphi})K]= -\mathrm{Tr}(\varphi K)+\frac{1}{2}\mathrm{Tr}(\varphi^2 K)
\end{equation}  
 \item \textit{p=2 }
 \begin{equation}
 -\frac{1}{2}\mathrm{Tr}[(1-e^{-\varphi})K]^2=-\frac{1}{2}\mathrm{Tr}(\varphi K \varphi K)
 \end{equation}
 \end{enumerate}
 Grouping the different terms in power of $\varphi$, we end up having
 \begin{equation}
 \label{cumulant_expansion}
 \begin{split}
 \log \mathbb{E}_K \left[ \exp\left(-\sum_{i=1}^\infty \varphi(a_i)\right)\right]=& -\mathrm{Tr}(\varphi K)+\frac{1}{2}\left[\mathrm{Tr}(\varphi^2 K)-\mathrm{Tr}(\varphi K \varphi K)\right]\\
 &+\text{higher order cumulants}
 \end{split}
 \end{equation}
Higher orders are given in the Appendix in Section \ref{appendix3} along with a conjecture relating the cumulant expansion to the Baker-Campbell-Hausdorff formula. Note that the successive terms in 
\eqref{cumulant_expansion} are also the cumulants of the quantity 
$\mathcal J=-\sum_{i=1}^\infty \varphi(a_i)$, i.e. 
$\mathbb{E}_K[\mathcal J^n]^c = \kappa_n(\varphi)$ defined in \eqref{defcum}.
\\

Consider now the determinantal processes related to the droplet and Brownian initial conditions. Evaluation of their cumulants in the large time limit allows in principle to
obtain the large deviation rate function for the left tail, 
\begin{equation}
\label{meerson11}
 \log \mathbb{E}_K \left[ \exp\left(-\sum_{i=1}^\infty \varphi(a_i)\right)\right] \underset{{t \to +\infty} }{\simeq}
 -t^2\Phi_-\left(\tilde s=\dfrac{s}{t^{2/3}}\right)
\end{equation}
in the limit $t \to +\infty$ at fixed negative $\tilde s$. In fact, in that limit one can replace 
in Eq.  \eqref{meerson11}
\begin{equation}
\label{max_replacement}
\varphi(a)=\log\left( 1+e^{t^{1/3}(a-s)} \right)\to \varphi_\infty(a) = t^{1/3}\max(0,a-s).
\end{equation}

We will now evaluate the lowest order cumulants in Eq. \eqref{cumulant_expansion}
replacing $\varphi(a)$ by $\varphi_\infty(a)$, for the determinantal processes related 
to the droplet and the Brownian initial conditions. For the droplet case, we will check that each cumulant brings one term in the rate function of Eq. 
\eqref{eq_Meerson}, obtained in Ref. \cite{sasorov2017large}, which we recall here in its 
integral version, i.e. 
\begin{equation}
\label{meerson22}
\Phi_-(\tilde{s})=\frac{1}{\pi^2} \int_{\tilde{s}}^0 \mathrm{d}v \; (v-\tilde{s})\left( \sqrt{1-\pi^2 v}-1\right)
\end{equation}
This formula is useful as it reminds of the trace operation applied to $\varphi_\infty$ and $K$.
Expanding this rate function for large negative argument, we seek to obtain 
\begin{equation}
\label{goalgoal}
\Phi_-(\tilde{s})=\frac{4 (-\tilde{s})^{5/2}}{15 \pi }-\frac{\tilde{s}^2}{2 \pi ^2}+\frac{2 (-\tilde{s})^{3/2}}{3 \pi ^3}+\frac{2 \tilde{s}}{3 \pi ^4}+\frac{\sqrt{-\tilde{s}}}{2 \pi
   ^5}-\frac{4}{15 \pi ^6}+O\left(\sqrt{\frac{1}{-\tilde{s}}}\right)
\end{equation}
In particular, we will see with our simple method that the leading term in $(-\tilde{s})^{5/2}$ is given by the first cumulant, the subleading term $\tilde{s}^2$ is given by the second cumulant and the next subleading term 
$(-\tilde{s})^{3/2}$ is given by the third cumulant. This claim has been confirmed analytically for the first cumulant and numerically for the second and third one. We conjecture that it holds to any order,
and furthermore that it extends to the Brownian IC kernel as well (although we have not yet an explicit
calculation for it).
\subsection{First cumulant}

By using the max replacement of Eq. \eqref{max_replacement}, we express the first cumulant as follows
\begin{equation} \label{dens1} 
-\mathrm{Tr}(\varphi_\infty K)=-t^{1/3}\int_{s}^{+\infty}\mathrm{d}v\,  (v-s) K(v,v)
\end{equation}
It is important to note that this first cumulant only involves the density of states of the determinantal process $K(v,v)=\rho(v)$. Indeed, as we will see in this Section, the tail of the large deviations 
is only governed by this density. In the case of the Airy process (droplet IC), it is 
well known that this density matches smoothly the square-root semi-circle bulk density of the GUE for
negative $v$ (see below) and decays as a stretched exponential for positive $v$. As the deformed Airy process (Brownian IC) shares the same asymptotic density of states as the Airy process as proved in the Appendix in Section \ref{details_calculation_sine}, we expect the large deviations for both IC to share the same tail.\\

Since we consider $s=\tilde{s} t^{2/3}$ large, the integral in Eq. \eqref{dens1} is dominated by values of order $v = O(t^{2/3})$, hence we can use the (bulk) asymptotic density of states for the Airy kernel
\be
K_{\mathrm{Ai}}(v,v)\underset{-v\gg1}{\simeq} \frac{\sqrt{\abs{v}}}{\pi}\Theta(-v)+O(\frac{1}{v})
\ee 
immediately leading to
\be  \label{int1} 
-\mathrm{Tr}(\varphi_\infty K)= -t^2 \frac{1}{\pi} \int_{\tilde{s}}^0 \mathrm{d}v \; (v-\tilde{s}) \sqrt{|v|}  
+O(t\tilde s)  = -\frac{4 t^2}{15\pi} (-\tilde s)^{5/2}+O(t\tilde s)
\ee
This quantity is exactly the leading order for the rate function that we were looking for 
in Eq. \eqref{goalgoal}. Note how simply it comes from the present calculation,
as the contribution of the bulk density of states. Note also that the {\it exact}, i.e. all orders,
result \eqref{meerson22} could be in principle obtained by the following replacement
in Eq.  \eqref{int1}
\be
\frac{\sqrt{\abs{v}}}{\pi}\Theta(-v) \to \frac{1}{\pi^2}  \left( \sqrt{1-\pi^2 v}-1\right) \Theta(-v) 
\ee 
i.e. a small modification of the density of states. We have however no explanation at this stage
for this observation. In the next Section we will obtain its series expansion from the higher
cumulants.

Before doing so, we note that if we want more control on the integral \eqref{dens1}, we can use the fact that for the Airy kernel, the first cumulant admits a closed form using the identity $K_{\mathrm{Ai}}(v,v)=\mathrm{Ai}'(v)^2-v\mathrm{Ai}(v)^2$.
\begin{equation}
\begin{split}
-\mathrm{Tr}(\varphi_\infty K)&=-\frac{t^{1/3}}{30} \left(\left(3-8 s^3\right) \text{Ai}(s)^2+8 s^2 \text{Ai}'(s)^2+4 s
   \text{Ai}(s) \text{Ai}'(s)\right)\\
   &=-\frac{t^{1/3}}{30} \left(8s^2 K_{\mathrm{Ai}}(s,s)+\mathrm{Ai}(s)(3\mathrm{Ai}(s)+4s\mathrm{Ai}'(s))\right)\end{split}
\end{equation}
For $-s\gg1$, the leading term is the first one
\begin{equation}
\label{first_cumulant_control}
-\mathrm{Tr}(\varphi_\infty K)=-\frac{4 t^{1/3}}{15} s^2 K_{\mathrm{Ai}}(s,s)+t^{1/3}O(s)
\end{equation}
leading to \eqref{int1}. We shall now consider higher cumulants to determine whether they give the successive orders of the rate function $\Phi_-$.
\subsection{Numerical confirmation of the second and third cumulant}
The expression to evaluate to obtain the second cumulant is the following, see Eq. \eqref{cumulant_expansion},
\begin{equation}
\begin{split}
&\frac{1}{2}\left[\mathrm{Tr}(\varphi_\infty^2 K)-\mathrm{Tr}(\varphi_\infty K \varphi_\infty  K)\right] = t^{2/3} \gamma_2(s) \\
& \gamma_2(s) = \frac{1}{2}\int_{s}^{+\infty}\mathrm{d}v\,  (v-s)^2K_{\mathrm{Ai}}(v,v)-
\frac{1}{2}\int_{s}^{+\infty}\int_{s}^{+\infty}\mathrm{d}a_1 \mathrm{d}a_2 (a_1- s)(a_2-s) K_{\mathrm{Ai}}(a_1,a_2)^2
\end{split}
\end{equation}
Although the first integral can be computed exactly using the closed formula for the Airy kernel or the approximation for the density of state, 
\begin{equation}
\frac{t^{2/3}}{2}\int_{s}^{+\infty}\mathrm{d}v\,  (v-s)^2K_{\mathrm{Ai}}(v,v)=\frac{8}{105\pi }t^3(-\tilde{s})^{7/2}+O(t(-\tilde{s})^{1/2}),
\end{equation}
the second integral, which is supposed to cancel the $t^3(-\tilde{s})^{7/2}$ term 
and give the $\tilde{s}^2$, remains a challenge.\\
~\\
Using the formulae for the second and third cumulant, see Eq. \eqref{third_order_expansion} in the Appendix in Section \ref{appendix3}, we have verified numerically on Mathematica that we actually find the first three terms of the series expansion of Ref. \cite{sasorov2017large} $\Phi_-$ function with the coefficients matching perfectly, that is we have checked with very good accuracy that
\be
\gamma_2(s) \underset{s \to -\infty}{\simeq} \frac{s^2}{2 \pi^2} \qquad , \qquad \gamma_3(s) \underset{s \to -\infty}{\simeq}  \frac{2 (-s)^{3/2}}{3 \pi^3} 
\ee
where $\gamma_3(s)$ is the third cumulant, see Eq. \eqref{third_order_expansion}
divided by $t$. We leave the analytic calculation of these cumulants to future work.
\subsection{General cumulants and discussion of the result from Ref. \cite{sasorov2017large}}
\label{scalingreasoning}
For the $n$-th cumulant, we will have a product of $n$ factors $t^{1/3} \max(0,a-s)$ giving an overall factor $t^{n/3}$ multiplied by a resulting integral $\gamma_n(s)$, depending only on $s$. To match the scaling function of Ref. \cite{sasorov2017large}, we need the leading order in terms of large negative $s$ to take a scaling form $t^2 \Phi(\frac{s}{t^{2/3}})$. Therefore, calling $c_n s^{\alpha_n}$ the leading term of 
$\gamma_n(s)$, and recalling that $s=\tilde s t^{2/3}$, we require the scaling identity
\begin{equation}
\label{scalingargument}
\alpha_n=3-\frac{n}{2}
\end{equation}
Nonetheless, looking at the large deviation rate function of Ref. \cite{sasorov2017large},
\begin{equation}
\Phi_-(\tilde{s})=\frac{4}{15\pi^6}(1-\pi^2 \tilde{s})^{5/2}-\frac{4}{15\pi^6}+\frac{2}{3\pi^4}\tilde{s}-\frac{1}{2\pi^2}\tilde{s}^2
\end{equation}
the last three terms seem to be anomalies. This implies, for the matching to work, that
the coefficients $c_n$ of all even cumulants for $n \geq 8$ vanish. The reason for such
a property remains to be understood.\\
~\\
The conclusion of this second method is that the $n$-th cumulant seems to give a contribution that is exactly equal to the $n$-th order expansion of Ref. \cite{sasorov2017large} rate function. This has been proved analytically for $n=1$, and numerically for $n=2$ and $n=3$. We conjecture that this remains true
for any $n$. 
%\newpage
\section{Extension of method B to earlier times }
\label{alltimes}
We will now discuss a possible extension of the previous calculations of the large deviation tail at large time to a tail valid at all times. If one does not proceed to any approximation to the function $\varphi$, then for both droplet and Brownian ICs having the same $\varphi$ function, the first cumulant of both  process with density $\rho$ is given exactly by the integral relation 
\begin{equation}
\kappa_1(t,s)=-\int_{\mathbb{R}}\mathrm{d}v\, \log(1+e^{t^{1/3}(v-s)}) \rho(v)
\end{equation}
\subsection{Extension to short time}
\label{extension_all_times}
It is interesting to note that the large time estimate of the Airy kernel, Eq. \eqref{KtAi}, is  quite reminiscent
of the short time estimate for the same kernels, which we now recall (for negative argument $\hat v<0$).
\begin{enumerate}
\item \textit{Droplet case : Airy kernel, see Ref. \cite{le2016exact} } 
\begin{equation}
K_{\rm Ai}(\frac{\hat v}{t^{1/3}},\frac{\hat v+t^{1/2}\kappa}{t^{1/3}})\underset{t\ll 1}{\simeq}\frac{1}{\pi t^{1/6}}\frac{\sin(\sqrt{-\hat v}\kappa)}{\kappa}
\end{equation}
\item \textit{Stationary case : deformed Airy kernel, see Ref. \cite{krajenbrink2017exact}} \\
\begin{equation}
\label{stat_short_time}
\begin{split}
&  K_{\rm Ai, \Gamma}(\frac{\hat v}{t^{1/3}},\frac{\hat v+t^{1/2}\kappa}{t^{1/3}})\underset{t\ll 1}{\simeq}\frac{1}{\pi t^{1/6}}\frac{\sin(f_{\tilde{w}}(\tilde v)\kappa)}{\kappa}\\
  &\tilde{v}=\hat v-\ln(w^2),\quad  \tilde{w}=wt^{1/2}, \quad f_{\tilde{w}}(\tilde{v})=\sqrt{W_0(\tilde{w}^2e^{-\tilde{v}+\tilde{w}^2})-\tilde{w}^2}
  \end{split}
\end{equation}
where $W_0$ is the first branch of the Lambert function.
\end{enumerate}
It seems that doing the transformation $t\mapsto t^{-1/2}$ allows to go from one regime to the other one, perhaps revealing a hidden symmetry in the behavior of the KPZ solution.\\

We apply method B to short time and evaluate the first cumulant $\kappa_1$ with the asymptotic kernels. For this, we proceed to the rescaling $v=\hat v t^{-1/3}$ and $s=\hat s t^{-1/3}$ and obtain 
\begin{equation}
\kappa_1(t,s=\frac{\hat s}{t^{1/3}})=-t^{-1/3}\int_{\mathbb{R}}\mathrm{d}\hat v\, \log(1+e^{\hat v-\hat s}) \rho(\hat v t^{-1/3})
\end{equation}
\begin{enumerate}
\item For the droplet IC, the density converges at small time to
\begin{equation} 
 \rho(\hat v t^{-1/3})\underset{t\ll 1}{\longrightarrow} \frac{t^{-1/6} }{\pi}\sqrt{\abs{\hat v}}\theta(-\hat v)
\end{equation}
which yields for the first cumulant
\begin{equation}
\label{droplet_short_time_100}
\begin{split}
\kappa_1(t,s=\frac{\hat s}{t^{1/3}})&\simeq -\frac{1}{\pi\sqrt{t}}\int_{\mathbb{R}^-}\mathrm{d}\hat v\, \log(1+e^{\hat v-\hat s}) \sqrt{\abs{\hat v}}\\
&\simeq  \frac{1}{\sqrt{4\pi t}}\mathrm{Li}_{5/2}(-e^{-\hat s})\\
\end{split}
\end{equation}
Remarquably, the exact short time distribution of the moment generating function derived in Ref. \cite{le2016exact} Eq. (20) for the droplet IC is fully encoded in its first cumulant. 

\item For the Brownian IC with drift $w$, the density converges at small time, using the conventions of Eq. \eqref{stat_short_time}, to
\begin{equation} \label{densitySTstat} 
\rho(\hat v t^{-1/3})\underset{t\ll 1}{\longrightarrow} \frac{t^{-1/6} }{\pi}\sqrt{W_0(\tilde{w}^2e^{-\tilde{v}+\tilde{w}^2})-\tilde{w}^2}\; \theta(-\hat v)
\end{equation}
which yields for the first cumulant
\begin{equation}
\begin{split}
\kappa_1(t,s=\frac{\hat s}{t^{1/3}})&=-\frac{1}{\pi \sqrt{t}}\int_{\mathbb{R}^+}\mathrm{d}\tilde v\, \log(1+\frac{e^{-\tilde v-\hat s}}{\tilde{w}^2}) \sqrt{W_0(\tilde{w}^2e^{\tilde{v}+\tilde{w}^2})-\tilde{w}^2}\\
&=-\frac{1}{\pi \sqrt{t}}\int_{\mathbb{R}^+}\mathrm{d}y\, \log(1+\frac{e^{-y-\hat s}}{y+\tilde{w}^2})\left[1+\frac{1}{y+\tilde{w}^2} \right]\sqrt{y}\\
\end{split}
\end{equation}
where we performed the change of variable $y=W_0(\tilde{w}^2e^{-\tilde{v}+\tilde{w}^2})-\tilde{w}^2$. The first cumulant then again gives the full moment generating function for the Brownian IC coinciding exactly with the one derived in \cite{krajenbrink2017exact} Eq. (16).
\end{enumerate}
~\\
In particular, these observations imply that an argument of measure concentration should be sufficient to explain why the first cumulant gives the entire large deviation rate functions at short time. This is encouraging in the sense that this method fully applies to the short time study and simplifies the summation of traces of the Fredholm determinant presented in Refs. \cite{le2016exact, krajenbrink2017exact} and recalled in method A.
This simpler method will be further developed in 
\cite{UsHalfSpace}.

\subsection{Extension to all times}
A crucial step in the calculation of method B was the approximation of $\varphi$ by $\varphi_\infty$. We claim that this approximation is valid as long as  $st^{1/3}$ is large and negative.
\begin{equation}
\varphi(v)=\log(1+e^{t^{1/3}(v-s)})\underset{-st^{1/3}\gg 1}{\longrightarrow} t^{1/3}\max(0,v-s)
\end{equation}
In order to determine the tail of the distribution of $H$ we also used the convergence of the moment generating function to the cumulative probability. We claim that this convergence remains true in the same regime where $st^{1/3}$ is large and negative. 
\begin{equation}
\mathbb{E}\left[\exp(-e^{H-st^{1/3}}) \right]\underset{-st^{1/3}\gg 1}{\longrightarrow } \mathbb{P}(H<st^{1/3})
\end{equation}
This can be seen by noting that one can rewrite
\begin{equation} \label{gumbel} 
\mathbb{E}\left[\exp(-e^{H-st^{1/3}}) \right]=\mathbb{P}(H+\text{Gumb}<st^{1/3})
\end{equation}
where Gumb is a random variable independent of $H$ with a unit Gumbel distribution. Hence, since Gumb is an order one random variable with an extremely fast decay in the left tail, the limit where $st^{1/3}\to -\infty$ corresponds to large negative values for $H$.
Note that the same identity \eqref{gumbel} holds for the Brownian IC replacing
$H \to H + \chi$ where now $H, \chi$ and Gumb are independent random variables
with the same conclusion since $\chi$ is also of order unity. \\

If we now work at intermediate times, we can fix the time $t$ and work in the limit $-s \ll1$.
The first cumulant is then given by
\begin{equation}
\kappa_1(t,s)=-t^{1/3}\int_{s}^{+\infty} \mathrm{d}v (v-s)\rho(v) 
\end{equation}
where $\rho$ is the density of states. For droplet and Brownian ICs, the density converges for large negative argument to an asymptotic value 
\begin{equation}
 \rho(v)\underset{-v\ll 1}{\longrightarrow} \rho_\infty(v)= \frac{1 }{\pi}\sqrt{\abs{ v}}\theta(-v)
\end{equation}

We can add and subtract the asymptotic density to obtain a suitable form for the large deviations, indeed
\begin{equation}
\kappa_1(t,s)=-t^{1/3}\left[\int_{s}^{+\infty} \mathrm{d}v (v-s)\left( \rho(v)-\rho_\infty(v) \right)+\int_s^{+\infty} \mathrm{d}v (v-s)\rho_\infty(v)\right]
\end{equation}
Replacing the asymptotic density, we obtain for droplet and Brownian ICs 
\begin{equation}
\begin{split}
\kappa_1(t,s)&=-t^{1/3}\left[\int_{0}^{+\infty} \mathrm{d}v\,  (v-s) \rho(v) + \int_{s}^0\mathrm{d}v\, (v-s)\left(\rho(v)-\frac{\sqrt{\abs{ v}}}{\pi}\right)+\frac{4}{15\pi}(-s)^{5/2}\right]
\end{split}
\end{equation}
One then needs to control the first two integrals to ensure that the last $(-s)^{5/2}$ term is the right dominant one, i.e. that the left tail of the large deviations is 
\begin{equation}
\label{arbitrary_time_regime}
-\log  \mathbb{P}(H<st^{1/3}) \simeq -\frac{4}{15\pi}t^{1/3} (-s)^{5/2}
\end{equation}
For the case of the droplet IC, this can be shown by an explicit analytical calculation using the closed form of the density $\rho(v)=\mathrm{Ai}'(v)^2-v\mathrm{Ai}(v)^2$ and as in Eq. \eqref{first_cumulant_control}, we observe that the correction is of order $O(s)$ thus establishing Eq. \eqref{arbitrary_time_regime} in the region $s\to -\infty$. 
%More generally, one would need to call again to a measure concentration property.
\\

The problem of determining how the next cumulants, i.e. second, third cumulants, influence the large deviations is still open at intermediate times and we again conjecture that they are subdominant compared to the first cumulant providing the $(-s)^{5/2}$ tail. This opens a new path in the cumulant expansion method to derive the left large deviations at all times.

%\newpage

\section{Extension of method B to a Pfaffian point process }
\label{half_space_extension}
\subsection{The half-space KPZ problem : known results }
We now consider the half-space KPZ problem with droplet IC where Eq. \eqref{eq:KPZ} is considered for $x \in \mathbb{R}^+$ and we add the additional Neumann boundary condition
\begin{equation}
\label{neumann}
\forall t>0, \quad \partial_x h(x,t)\mid_{x=0}\, =A,
\end{equation}
which physically corresponds to the presence of a wall at $x=0$. 
\begin{comment}
The interest in this model lies in the context of constrained fluctuations with application to Casimir forces and extreme value statistics \cite{krech1994casimir, le2009fluctuation}. 
\end{comment}
Exact solutions have been obtained in a few cases. For $A=+\infty$, which corresponds to an absorbing wall if one represents the problem in terms of a directed polymer, see Ref. \cite{gueudre2012directed}, the large time limit corresponds to the statistics of the GSE random matrix ensemble. For $A=-\frac{1}{2}$, which corresponds to the critical case, see Ref. \cite{barraquand2017stochastic}, the large time statistics is described by the GOE ensemble. For the simplicity of the derivation, we now focus the critical case $A=-\frac{1}{2}$, other cases ($A=0$ and $A=+\infty$)
will be treated in \cite{UsHalfSpace}. For the critical case, it has been proved in \cite{barraquand2017stochastic} 
that (in our system of units) 
\begin{equation}
\label{halfspace1}
\mathbb{E}_{\mathrm{KPZ},\,   \mathrm{half-space}}\left[ \exp(-\frac{1}{4}e^{H(t)-st^{1/3}})\right]=\mathbb{E}_{\mathrm{GOE}}\left[ \prod_{i=1}^\infty\frac{1}{\sqrt{1+e^{t^{1/3}(a_i-s)}}}\right]
\end{equation}
where the set $\lbrace a_i \rbrace$ forms a GOE point process. Note that there is an extra factor $\frac{1}{4}$ in Eq. \eqref{halfspace1} compared to Eq. \eqref{fredholm1}. The reason for this if two-fold. The initial condition for the droplet case in the full-space was $h_{\mathrm{drop}}(x,0)=-\frac{\abs{x}}{\delta}-\ln(2\delta)$ with $\delta \ll 1$, in the half-space the normalization should be different and include a factor $\ln(\delta)$ instead of $\ln(2\delta)$ which accounts for a factor $\frac{1}{2}$ in the moment generating function.
The other $\frac{1}{2}$ factor can be seen as a square root in the first exponential
% indeed we have 
%\begin{equation}
%\mathbb{E}_{\mathrm{KPZ},\,  \mathrm{half-space}}\left[ \exp(-\frac{1}{4}e^{H(t)-st^{1/3}})\right]=\mathbb{E}_{\mathrm{KPZ},\,  \mathrm{half-space}}\left[ \sqrt{\exp(-\frac{1}{2}e^{H(t)-st^{1/3}})}\right],
%\end{equation} 
which corresponds to the square root in the expectation value over the GOE process \cite{footnote_barraquand}.

\subsection{Estimates at large time}
To study the large time limit, we again use method B of Section \ref{method2} and we define the quantities
\begin{equation}
\begin{split}
&\sigma _{t,s}(v)=\frac{1}{\sqrt{1+e^{t^{1/3}(v-s)}}}-1\\
& \varphi(v)=- 2 \log(1+\sigma_{t,s}(v))\underset{t\gg1}{\to} \varphi_\infty(v)=t^{1/3}\max(0,v-s)
\end{split}
\end{equation}
i.e. the function $\varphi(v)$ is the same as in the previous section. 
The GOE point process being a Pfaffian and not a determinantal point process, the formula analogous to Eq. \eqref{deter_process} involves a Fredholm Pfaffian, see Ref. \cite{rains2000correlation}
\begin{equation}
\label{halfspace2}
\mathbb{E}_{\mathrm{GOE}}\left[ \prod_{i=1}^\infty\frac{1}{\sqrt{1+e^{t^{1/3}(a_i-s)}}}\right]=\mathrm{Pf}(J+\sigma_{t,s} K_{\mathrm{GOE}})
\end{equation}
together with
\begin{equation}
J=\bigg(\begin{array}{cc}
0 & 1 \\ 
-1 & 0
\end{array} 
\bigg) \qquad K_{\mathrm{GOE}}= \bigg(\begin{array}{cc}
K_{11} & K_{12} \\ 
-K_{12} & K_{22}
\end{array} 
\bigg) 
\end{equation}
The off-diagonal element is defined, see Ref. \cite{barraquand2017stochastic}, as 
\begin{equation}
\label{GOE_offdiagonal1}
K_{12}(x,y)=\frac{1}{(2i\pi)^2}\mkern-10mu\int\limits_{-i\infty+ \epsilon}^{+i\infty+ \epsilon}\int\limits_{-i\infty}^{+i\infty} \mathrm{d}w \mathrm{d}z \frac{w-z}{2w(w+z)}e^{\frac{z^3}{3}+\frac{w^3}{3}-xz-yw} + \frac{1}{2} \mathrm{Ai}(x)
\end{equation}
where $\epsilon=0^+$ which is also
\be \label{GOE_Airy}
K_{12}(x,y)= K_{\rm Ai}(x,y) - \frac{1}{2} {\rm Ai}(x) \int_0^{+\infty} dr {\rm Ai}(y+r) 
+ \frac{1}{2}  {\rm Ai}(x)
\ee 
The key relation to use method B of Section \ref{method2} is 
\begin{equation}
\label{halfspace3}
\mathrm{Pf}(J+\sigma_{t,s} K_{\mathrm{GOE}})^2 =\mathrm{Det}(I-\sigma_{t,s} J K_{\mathrm{GOE}})
\end{equation}
We recall at late time the convergence of the moment generating function to the cumulative probability
\begin{equation}
\label{halfspace4}
\mathbb{E}_{\mathrm{KPZ}}\left[ \exp(-\frac{1}{4}e^{H(t)-st^{1/3}})\right]\underset{t\gg 1, \,  s<0}{\longrightarrow} \mathbb{P}(H(t)<st^{1/3})
\end{equation}
So, at large time, combining Eqs. \eqref{halfspace1}, \eqref{halfspace2}, \eqref{halfspace3} and \eqref{halfspace4}, we obtain  
\begin{equation}
\label{halfspace5}
\mathbb{P}(H(t)<st^{1/3})^2 \underset{t\gg 1}{\simeq}  \mathrm{Det}(I-\sigma_{t,s} J K_{\mathrm{GOE}})\underset{t\gg 1}{\simeq} \mathrm{Det}(I-[e^{-\varphi_\infty/2} -1]  JK_{\mathrm{GOE}})
\end{equation}
Taking the logarithm of Eq. \eqref{halfspace5}, we can now apply the method B of Section \ref{method2} to calculate the tail of the large deviation rate function using the first cumulant of the Pfaffian point process 
\begin{equation}
\begin{split}
2\log \mathbb{P}(H(t)<st^{1/3})&\simeq \frac{1}{2} \mathrm{Tr}(\varphi_\infty  JK_{\mathrm{GOE}})\\
&\simeq-\; \mathrm{Tr}(\varphi_\infty K_{12})\\
&\simeq-t^{1/3}\int_{s}^{\infty} \mathrm{d}v\, (v-s)K_{12}(v,v)
\end{split} 
\end{equation}
For large negative argument we prove in the Appendix in Section \ref{GOE_asymptotic} the following asymptotics expression for the off-diagonal element of the GOE kernel is the edge of the semi-circle density
\begin{equation}
K_{12}( v ,  v )\underset{-v\gg 1}{\simeq} \frac{\sqrt{\abs{v}}}{\pi}\theta(-v)
\end{equation}
which leads, after redefining $s=\tilde{s}t^{2/3}$, to
\begin{equation}
\begin{split}
\log \mathbb{P}(H(t)<st^{1/3})\simeq  -\frac{2t^2}{15\pi} (-\tilde{s})^{5/2}\\
\end{split}
\end{equation}
By the same scaling reasoning of Section \ref{scalingreasoning} and the same scaling argument of Eq. \eqref{scalingargument}, higher order cumulants cannot yield a $(-\tilde{s})^{5/2}$ contribution, therefore we are ensured to obtain the correct expression for the large deviation tail, which is more tedious to see for the Pfaffian process than for the determinantal process due to the presence of diagonal and off-diagonal elements in the kernel. We observe that \textit{the large deviation tail for the half-space is half of the one of the full-space}, i.e. $c_{\mathrm{HS}}=\frac{2}{15\pi}$ instead of $c_{\mathrm{FS}}=\frac{4}{15\pi}$.\\

This fact can be understood by examining the case $A=+\infty$ where the system is effectively cut into two independant half spaces. As stated in Ref. \cite{gueudre2012directed} Eq. (32), in that case we have the inequality
\begin{equation} \label{twohalfves} 
\mathbb{E}_{\mathrm{KPZ}, \, \mathrm{full-space}}\left[ \exp(-e^{H(t)-st^{1/3}})\right]<\mathbb{E}_{\mathrm{KPZ},\, \mathrm{half-space}}\left[ \exp(-e^{H(t)-st^{1/3}})\right]^2 
\end{equation}
where both expectation values are taken over the droplet initial condition. This inequality indeed implies that the coefficient of the tail of the full-space is at least twice the one of the half-space, i.e. $2 c_{\mathrm{HS}}\leq c_{\mathrm{FS}}$. In the critical case $A=-\frac{1}{2}$, we have shown that $2 c_{\mathrm{HS}}= c_{\mathrm{FS}}$ and it is reasonable to expect that this still holds for arbitrary value of $A\geq -\frac{1}{2}$ although this remains to be checked. We have therefore obtained the tail of the large deviation rate function for a half-space problem and as in Section \ref{alltimes} we expect this to be valid for all times and sufficiently large negative $H$.\\

Under the assumption that the short time behavior of the moment generating function can be entirely captured through the first cumulant of the Pfaffian point process, we will derive in a future work 
\cite{UsHalfSpace} 
the exact short time height distribution for the half-space problem.

\section{Exact bounds}
It is possible, using the method B, to obtain exact bounds on the large deviations of the KPZ solution.
\subsection{First cumulant bound}
 The simplest one comes from the first cumulant $\kappa_1(\varphi)$ using Jensen's inequality $
\mathbb{E}_K[\exp \mathcal J] \geq \exp \mathbb{E}_K[\mathcal J]$ for $\mathcal J=-\sum_{i=1}^\infty \varphi(a_i)$, we obtain for a set $\lbrace {a}_i \rbrace $ following a determinantal process with kernel $K$, a bound for the averages introduced in \eqref{deter_process}
\be
- \log \mathbb{E}_{K}\left[ \prod_{i=1}^{\infty} \frac{1}{1+e^{t^{1/3}(a_i-s)}}\right] 
\leq -\kappa_1(\varphi) =  \int_{\mathbb{R}} \mathrm{d}v  \log( 1 + e^{t^{1/3}(v-s)}) K(v,v)
\ee
Specializing to $K=K_{\rm Ai}$ one obtains an exact bound for the following average
over the KPZ noise for droplet initial condition (see \eqref{KPZdet} and \eqref{gumbel})
\bea
\!\!\!\!\!\!\!\!\!\! \!\!\!\!\!\! - \log \mathbb{E}_{\, \mathrm{KPZ}}\left[ \exp\left(  -e^{H(t)-st^{1/3}}\right) \right]
&=& - \log \mathbb{P}(H(t)+\text{Gumb}<st^{1/3}) \\
&\leq& -\kappa_1(\varphi) \\
&=&  \int_{\mathbb{R}} \mathrm{d}v  \log( 1 + e^{t^{1/3}(v-s)})K_{\rm Ai}(v,v) \nonumber
\eea
valid for all times and $s$. At short time we have observed that this bound is actually saturated to
leading order in $t$. In the long time limit it leads to an upper bound on the left tail large deviation rate function $\Phi_-(z)$. Indeed, using the bounds $ \log(1+e^x)\leq \max(0,x)+\log 2$ and $\log(1+e^x)\leq e^x$, and writing the density $K_{\mathrm{Ai}}(v,v)=\rho(v)$ one has
\begin{equation}
\label{right_bound}
\begin{split}
&- \log \mathbb{P}(H(t)+\text{Gumb}<z t)
%- \log \mathbb{E}_{\, \mathrm{KPZ}}\left[ \exp\left(  -e^{H(t)-st^{1/3}}\right) \right]
\\
&\leq t^{1/3} \int_{s}^{+\infty}\mathrm{d}v (v-s)\rho(v)+\log 2\int_{s}^{+\infty}\mathrm{d}v \rho(v)+\int_{-\infty}^s \mathrm{d}v e^{t^{1/3}(v-s)}\rho(v)\\
&\leq t^2 \int_{z}^{+\infty}\mathrm{d}\tilde v (\tilde v-z)\frac{\rho(t^{2/3}\tilde v)}{t^{1/3}}+t\log 2\int_{z}^{+\infty}\mathrm{d}\tilde v \frac{\rho(t^{2/3}\tilde v)}{t^{1/3}}+t\int_{-\infty}^z \mathrm{d}\tilde v e^{t(\tilde v-z)}\frac{\rho(t^{2/3}\tilde v)}{t^{1/3}}\\
\end{split}
\end{equation}
with $z= s t^{-2/3}$.
At large time, it was shown that the ratio $\rho(t^{2/3}\tilde v)/t^{1/3}$ admits a non-trivial limit which is the edge of Wigner's semi-circle. Besides, the large parameter governing the left large deviations is $t^2$ as seen in \eqref{meerson11}, we therefore take the large time limit and obtain an upper bound on the large deviation rate function
\begin{equation}
\begin{split}
\Phi_-(z)=-\lim_{t\to +\infty} \frac{1}{t^2} \log \mathbb{P}(H(t)<z t)
%\log \mathbb{E}_{\, \mathrm{KPZ}}\left[ \exp\left(  -e^{H(t)-zt}\right) \right] 
&\leq  \int_{z}^{+\infty}\mathrm{d}\tilde v (\tilde v-z)\frac{\sqrt{|\tilde v|}}{\pi}\theta(-\tilde v)\\
& \leq \frac{4}{15\pi} |z|^{5/2}  \theta(-z) \label{boundfirst} 
\end{split}
\end{equation}
where the unit Gumbel random variable can be neglected in that limit.
This bound is saturated for large negative $z$ as shown in this paper. 
Note that this bound is sufficient to exclude the earlier conjecture $\Phi_-(z)=\frac{1}{12} z^3 \theta(-z)$
of \cite{LargeDevUs}. If $z\geq 0$, then the r.h.s gives 0 which confirms that the right large deviations should be on a different scale than $t^2$.

This bound can be applied to the KPZ equation with other initial conditions. For the Brownian initial condition the kernel $K$ is the (time-dependent) deformed Airy kernel $K_{\rm Ai, \Gamma}$ given in \eqref{deformedK} 
and the density is $\rho_{\rm Br}(v)=K_{\rm Ai, \Gamma}(v,v)$. The asymptotic formula for the
density at short time is given in \eqref{densitySTstat}. At large time is equals the one for the droplet IC, $\rho_{\rm Br}(v) \simeq \frac{1}{\pi} \sqrt{|v|} \theta(-v)$, as seen from \eqref{KtAi}. Hence we have for arbitrary time
\be
 - \log \mathbb{P}(H(t)+\text{Gumb} + \chi <st^{1/3}) \leq 
  \int_{\mathbb{R}} \mathrm{d}v  \log( 1 + e^{t^{1/3}(v-s)}) \rho_{\rm Br}(v)
\ee
and in the large time limit the same bound \eqref{boundfirst} for the Brownian initial condition.

For the critical half-space KPZ, using the Jensen inequality with
$\mathcal J=- \frac{1}{2} \sum_{i=1}^\infty \varphi(a_i)$ and the average over the
GOE process, we similarly obtain the bound 
\be
- \log \mathbb{P}(H(t) - \log 2 +\text{Gumb}<st^{1/3}) \leq \frac{1}{2}
 \int_{\mathbb{R}} \mathrm{d}v  \log( 1 + e^{t^{1/3}(v-s)}) \rho_{\rm GOE}(v) \nonumber
\ee
where 
\be
\rho_{\rm GOE}(v) = K_{12}(v,v) 
= {\rm Ai'}(v)^2-v  {\rm Ai}(v)^2 + \frac{1}{2} {\rm Ai}(v) \int_{-\infty}^v dr {\rm Ai}(r) 
\ee
In the large time limit this shows, similarly as above 
$\Phi_-(z) \leq \frac{2}{15\pi} |z|^{5/2}  \theta(-z)$.

\subsection{Conditional bound}

We now sketch an improvement on the above bound based on conditioning the
determinantal point process $a_i$ on the value of its maximum, $a_1 = \max_i a_i$. This improvement is also discussed in \cite{corwin2018lower} and \cite{AllofUs}. Let us define the conditional density 
\be \label{conddens} 
\rho_b(a)=\mathbb{E}\left[ \sum_{i=1}^{+\infty} \delta(b-a_i)\mid a_1=a \right]
\ee
which by definition vanishes for $b>a$. We can use the conditional Jensen inequality (see e.g. \cite{durrett2010probability}) for the quantity $\mathcal{J}= - \sum_i \varphi(a_i)$ and obtain
\bea
&& \mathbb{E}_K[ e^\mathcal{J} ] = \int da P(a_1=a) \mathbb{E}_K[ e^\mathcal{J} |a ]
\leq  \int da P(a_1=a) e^{ \mathbb{E}_K[ \mathcal{J} |a ] } \\
&& = \int da \exp\left( \ln P(a_1=a)  - \int_{-\infty}^a db \, \rho_a(b) \ln(1 + e^{t^{1/3} (b-s)}) \right)
\label{exp}
\eea
where we use the loose notation $P(a_1=a)$ for the PDF of $a_1$. This inequality is
true for all time (and for any determinantal process). Let us apply it now to the large time
limit of the droplet initial condition, i.e. $K=K_{\rm Ai}$
and $P(a_1=a)= F_2'(a)$ the PDF of the GUE Tracy Widom distribution. We now define the rescaled variables
$a= t^{2/3} \tilde a$, $b= t^{2/3} \tilde b$ and $s= t^{2/3} \tilde z$, and use the well known left tail of the Tracy-Widom distribution 
$P(a_1=t^{2/3} \tilde a) \simeq_{t \to +\infty} \exp(- t^2 \frac{|\tilde a|^3}{12})$. Assuming the existence of the limit
\be
\tilde \rho_{\tilde a}(\tilde b) := \lim_{t \to +\infty} t^{-1/3}  \rho_{t^{2/3} \tilde a}(t^{2/3} \tilde b) 
\ee 
one sees that all the terms in the exponential in \eqref{exp} are uniformly of $O(t^2)$, hence one can apply the
saddle point method and obtain the following upper bound for the large deviation rate function
(defined above)
\begin{equation} \label{minimiz}
\Phi_-(z) \leq  \Phi^{\rm sup}_-(z) = \min_{\tilde a \leq 0} \left[\frac{1}{12} |\tilde {a}|^3 
+\theta(\tilde a-z) \int_{z}^{\tilde{a}} \mathrm{d}\tilde b \, (\tilde b-z) \tilde \rho_{\tilde a}(\tilde b)\right]
\end{equation}

We now consider the limit of the scaled conditional density $\tilde \rho$ and infer it from the 
work of Perret and Schehr \cite{perret2014near,perret2015density}. We obtain (droping the
tilde on the variables)
\be \label{densscaled}
\tilde \rho_a(b) = \frac{a- 2b}{2 \pi \sqrt{a-b}} \theta(a-b)
\ee 
This is found as follows. The conditional density that we introduced 
above in \eqref{conddens} is related to the quantity $\tilde \rho_{\rm edge}$
calculated in \cite{perret2015density}, see Eq. (31)-(33) there, as
follows
\be
\tilde \rho_a(b) = \tilde \rho_{\rm edge}(a-b|a) 
\ee
We thus need the behavior of $\tilde \rho_{\rm edge}(\tilde r|x)$ 
when both $\tilde r \to +\infty$ and $x \to -\infty$ at the same rate.
We can use tentatively the asymptotics Eq. (80) of \cite{perret2014near}
\begin{equation}
\begin{split}
&\tilde{f}(\tilde r,s)\sim_{r\to +\infty} 2^{-1/6} \frac{1}{\tilde r^{1/4}}\sin(\frac{2}{3}\tilde r^{3/2} - x \sqrt{\tilde r}  +\frac{\pi}{4}) + O(\tilde r^{-3/4} ) \\
&\tilde{g}(\tilde r,s)\sim_{r\to +\infty} 2^{-1/6} \tilde r^{1/4} \cos(\frac{2}{3} \tilde r^{3/2} - x \sqrt{\tilde r} +\frac{\pi}{4}) + O(\tilde r^{-1/4} )  
\end{split}
\end{equation}
as well as $q(x) \simeq_{x\to -\infty} \sqrt{\frac{-x}{2}}$ and
$R(a)\simeq_{x\to -\infty} \frac{x^2}{4}$. 
We note from the presence of the factor ${\cal F}_2(x)$,
that the formula (33) in \cite{perret2015density} is presumably for the joint
density and not the conditional one. Hence we must divide it
by ${\cal F}_2'(x) = R(x) {\cal F}_2(x)$. This then leads to 
our result \eqref{densscaled}. Note that these results can also
be obtained in principle by studying the resolvant (see Appendix V in Section \ref{app:conditional}).

One can also obtain \eqref{densscaled} from taking the edge scaling limit of
the formula obtained in \cite{dean2006large,dean2008extreme} for the same conditional
density but defined in the bulk of the GUE spectrum. There this density
is obtained by a saddle point method using a Coulomb gas approach.
The usual edge scaling states that $\sqrt{2} N^{1/6} (\lambda_i - \sqrt{2 N}) \to a_i$
in the sense of determinantal point processes. We must thus make the correspondence
between the variables $z,x$ of \cite{dean2006large} and our variables as
$z=- \sqrt{2} - \frac{a}{\sqrt{2} N^{2/3}}$ and
$x=\frac{a-b}{\sqrt{2} N^{2/3}}$. We then obtain from their
Eqs. (15-16) for the scaled density $f(x)$
\be \label{rhoab} 
\tilde \rho_a(b) = \lim_{N \to +\infty} \frac{N^{1/3}}{\sqrt{2}} f(x) = \frac{a- 2b}{2 \pi \sqrt{a-b}} \theta(a-b)
\ee 
which coincides with \eqref{densscaled}
(the prefactor in \eqref{rhoab} comes from the identification of the densities, using the notations of \cite{dean2006large},
$\frac{1}{N} \tilde \rho_a(b) \equiv \rho_N(\lambda) d\lambda = f(x) dx = f(x) \frac{db}{\sqrt{2} N^{2/3}}$).
The fact that the two calculations give the same result is a good indication
that our limit procedure is correct.

\begin{figure}[h!]
\centering
        \includegraphics[scale=0.65]{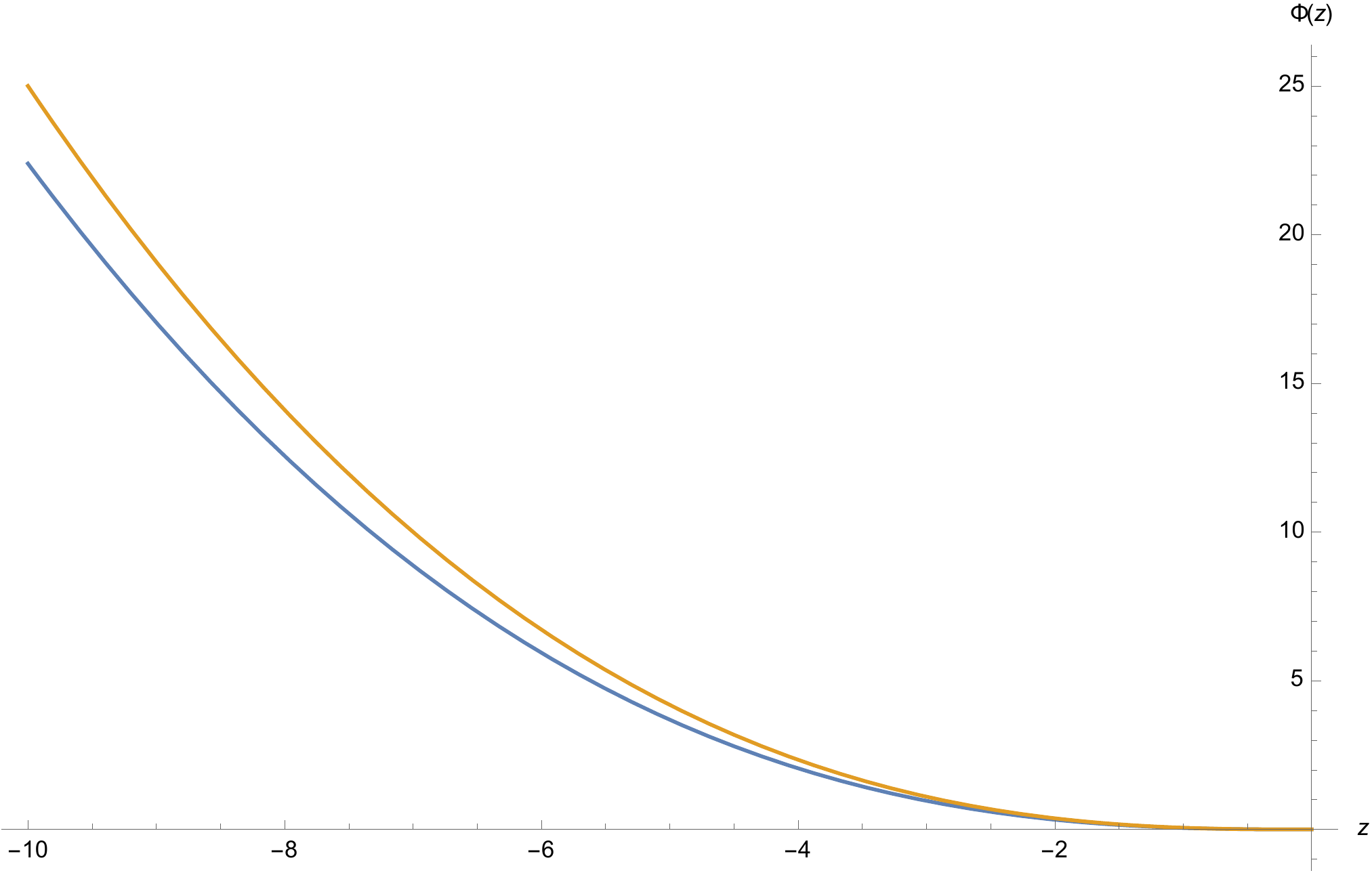}
        \caption{Comparison of the predicted rate function and its upper bound as a function of $z$.
        Orange: upper bound. Blue: predicted.}
                \label{fig1} 
\end{figure}

\begin{figure}[h!]
\centering
        \includegraphics[scale=0.65]{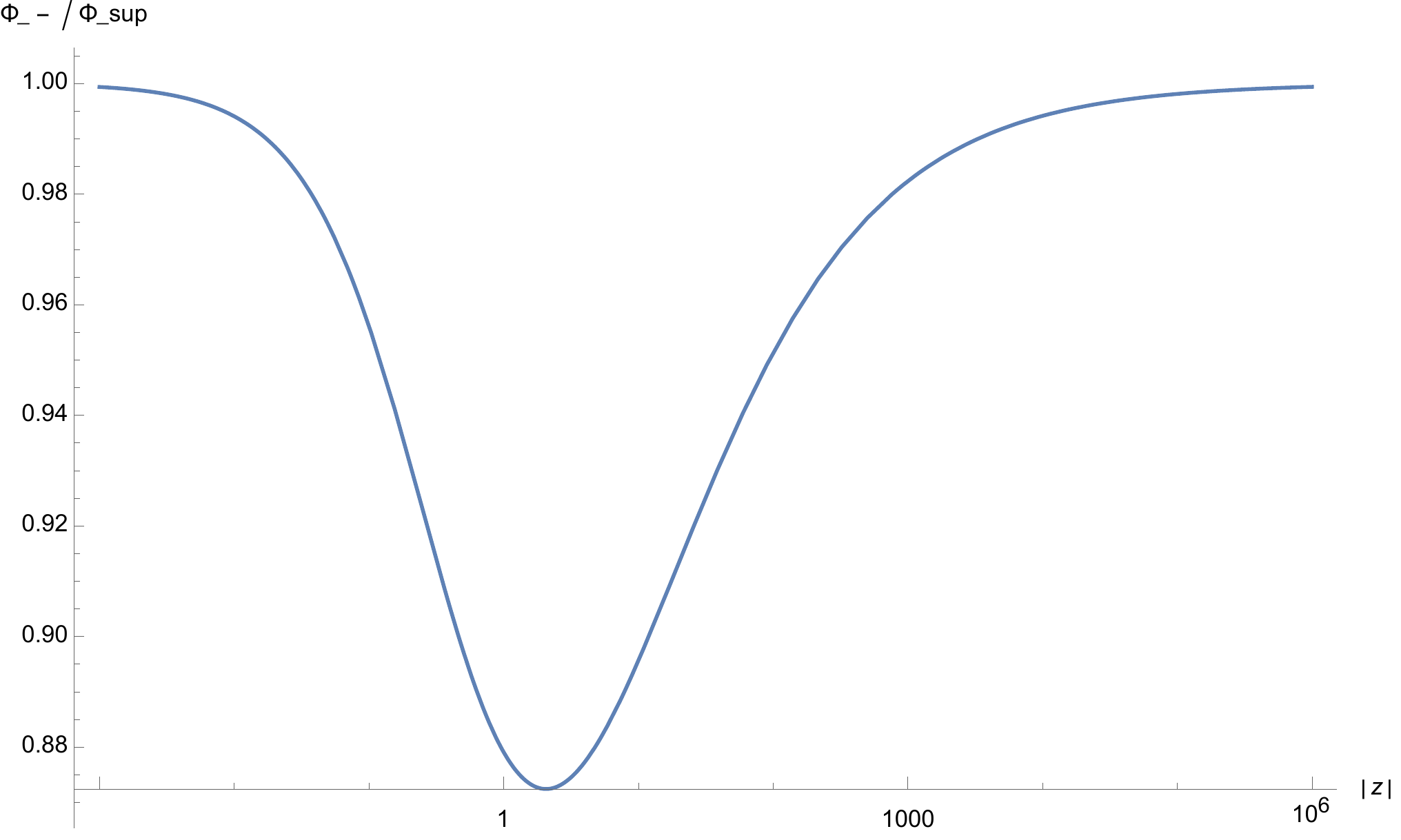}
        \caption{Ratio of the predicted rate function to its upper bound, $\Phi_-(z)/\Phi^{\rm sup}_-(z)$,
        as a function of $-z$ in log scale.}
                \label{fig2} 
\end{figure}

Inserting now the formula \eqref{densscaled}
for $\tilde \rho$ inside the minimization problem 
\eqref{minimiz}, leads to
\be
\Phi^{\rm sup}_-(z) = \min_{\tilde a \leq 0} \left[\frac{1}{12} |\tilde {a}|^3 
- \theta(\tilde a-z)  \frac{2 (\tilde a-z)^{3/2} (3 \tilde a+2 z)}{15 \pi } \right]
\ee
The minimum is attained for $z<\tilde a<0$ at the value $\tilde a=a^*=- \frac{4}{\pi^2} (\sqrt{4-\pi^2 z}-2)
= \frac{-z}{2 + \sqrt{4-\pi^2 z}}$. Computing the value at the minimum, one obtains the upper bound
$ \Phi_-(z) \leq  \Phi^{\rm sup}_-(z)$ with
\be
 \Phi^{\rm sup}_-(z) = 
%&& 
%\frac{320 \left(\sqrt{4-\pi ^2 z}-2\right)^3-8 \left(2 \pi ^2 z-12
%\sqrt{4-\pi ^2 z}+24\right) \left(-\pi ^2 z-4
%  \sqrt{4-\pi ^2 z}+8\right)^{3/2}}{60 \pi ^6} \\
\frac{16 (\sqrt{4-\pi ^2 z}-2)^3}{3 \pi ^6}-\frac{4 (\pi
^2 z-6 \sqrt{4-\pi ^2 z}+12)
  (-\pi ^2 z-4 \sqrt{4-\pi ^2 z}+8)^{3/2}}{15 \pi ^6}
\ee
One can compare the upper bound with the conjecture of
\cite{sasorov2017large}, denoted here $\Phi_-(z)$ and given in \eqref{eq_Meerson}
\begin{enumerate}
\item At small negative $z$ one has
\be
\Phi^{\rm sup}_-(z) =  \frac{|z|^3}{12}-\frac{\pi ^2 z^4}{192}+O\left(z^5\right) 
\quad , \quad   \Phi_-(z) = \frac{|z|^3}{12}-\frac{\pi ^2 z^4}{96}+O\left(z^5\right)
\ee
\item At large negative $z$ one has
\be
\Phi^{\rm sup}_-(z) =  \frac{4 |z|^{5/2}}{15 \pi }-\frac{8 |z|^{3/2}}{3 \pi ^3}
%+\frac{32 z}{3 \pi^4}+O\left(\frac{1}{\sqrt{\frac{1}{z}}}\right) 
+ O(|z|) \quad , \quad   \Phi_-(z) = \frac{4 |z|^{5/2}}{15 \pi }-\frac{z^2}{2 \pi ^2}+\frac{2 |z|^{3/2}}{3 \pi ^3}
%-\frac{2 z}{3   \pi ^4}+O\left(\frac{1}{\sqrt{\frac{1}{z}}}\right)
+ O(|z|)
\ee
\end{enumerate} 
The result of \cite{sasorov2017large} passes the test of the bound,
which is quite good, as can also be seen on the Figs. \ref{fig1} and \ref{fig2}.

Finally we note that a conditional bound in the case of the critical half-space 
can be deduced immediately from the above results.
Indeed, in Ref. \cite{dean2006large} it is shown that the conditional density is the same for GUE and GOE.
Hence we expect that $\tilde \rho_b(a)$ is the same function for the GOE edge point process 
and the Airy point process. On the other hand for the GOE case $P(a_1=a) \sim e^{- \frac{1}{24} |a|^3}$.
Since in that case the bound uses $\mathcal{J}= \frac{1}{2} \sum_{i} \varphi(a_i)$ (see previous subsection)
we obtain for the critical case $A=-1/2$
\be
\Phi^{\rm half-space}_-(z) \leq \frac{1}{2} \Phi^{\rm sup}_-(z) 
\ee
where $\Phi^{\rm sup}_-(z)$ is the upper bound for the full-space rate function given above.
Recall from Eq. \eqref{twohalfves} that for $A=+\infty$ one can show that
$\Phi^{\rm half-space}_-(z) \leq \frac{1}{2} \Phi^{\rm full-space}_-(z)$. In addition
from matching with the left tail of the GSE Tracy Widom distribution for the typical fluctuations found in 
\cite{gueudre2012directed},
one can surmise that for small negative $z$, $\Phi^{\rm half-space}_-(z) \simeq \frac{1}{24} |z|^3$ for $A=+\infty$.
Hence it is overall tempting to conjecture that $\Phi^{\rm half-space}_-(z) = \frac{1}{2} \Phi^{\rm full-space}_-(z)$ for all boundary conditions $A \geq -1/2$.

\section{Interpretation of the large deviations in terms of fermions}
\label{fermion_large_dev}
There exists an interpretation of the large deviations of the solution of the KPZ equation in terms of fermions in quantum mechanics. This section closely follows the one of Ref. \cite{le2016exact} where the short time large deviations studied there, referring to the high (reduced) temperature limit, is replaced here by the large time large deviation, now referring to the low (reduced) temperature limit of the quantum problem.\\

Consider the quantum problem of $N$ non-interacting spinless fermions of mass $m$ in an harmonic trap at finite temperature $T$ described by the Hamiltonian $H=\sum_{i=1}^N\frac{p_i^2}{2m}+\frac{1}{2}m\omega^2 x_i^2$. We use $x^*=\sqrt{\hbar /m\omega}$ and $T^*=\hbar \omega$ as units of length and energy. At $T=0$, in the ground state and for large $N$, the average fermion density is given by the semi-circle law with an edge at $x_{\mathrm{edge}}=\sqrt{2N}$. At finite temperature, the behavior of the physical quantities in the bulk changes on a temperature scale $T\sim N$ (bulk scaling), while near the edge it varies on a scale $T=N^{1/3}/b$, where $b$ is the {\it inverse reduced temperature}, a parameter of order one. Here we are interested in the position of the rightmost fermion $x_{\mathrm{max}}(T)$. Its cumulative distribution was shown, see Refs. \cite{dean2015finite, dean2016noninteracting}, to be given by the same Fredholm determinant as in Eq. \eqref{fredholm1} for the solution of the KPZ equation with droplet IC.
\begin{equation}
\mathbb{P}\left(\frac{x_{\mathrm{max}}(T)-x_{\mathrm{edge}}}{w_N}\leq s\right) \simeq Q_{t=b^3}(s)
\end{equation}
where $w_N=N^{-1/6} /\sqrt{2}$, and the equivalence is valid for $N, T \to +\infty$ with $b$ fixed.

Since we have already analyzed the long time limit $t \gg 1$ of the Fredholm determinant $Q_t(s)$, this provides us with an explicit formula for the fermion problem valid in the low reduced temperature region $b \gg 1$ of the edge scaling regime. For this aim, we apply the scaling $s=\tilde{s}t^{2/3}$, where $t^{2/3}=N^{2/3}/T^2$. This lead us to define $w_N(T)= t^{2/3}w_N = b^2 w_N=\sqrt{\frac{N}{2}}\frac{1}{T^2}$. Thus $w_N(T)$ sets the scale of fluctuations of $x_{\mathrm{max}}$ for the regime $t\gg 1$ corresponding to $T\ll N^{1/3}$. \\

Using the large deviation of the Fredholm determinant for large negative argument, we observe the relation in terms of fermions
\begin{equation}
\log \mathbb{P}\left(\frac{x_{\mathrm{max}}(T)-x_{\mathrm{edge}}}{w_N(T)}\leq \tilde{s }\right) \simeq -\frac{4}{15\pi}\frac{N^2}{T^6} (-\tilde{s})^{5/2} \label{leftlarge} 
\end{equation}
valid for $\tilde s$ fixed but large negative, and $b$ fixed. More generally, given the large deviation form of Eq. 
\eqref{left} we expect that
\begin{equation}
\log \mathbb{P}\left(\frac{x_{\mathrm{max}}(T)-x_{\mathrm{edge}}}{w_N(T)}\leq \tilde{s }\right) \simeq - b^6 \tilde \Phi_-(\tilde s=\frac{s}{b^2}) \label{sb2} 
\end{equation}
which reduces to Eq. \eqref{leftlarge} for large negative $\tilde s$. In the opposite limit of $\tilde s \to 0^-$, i.e.
$b \to +\infty$,
using $\tilde \Phi_-(\tilde s) \simeq \frac{1}{12} \abs{ \tilde s} ^3 $,
see Refs. \cite{dean2016noninteracting,LargeDevUs}, 
one matches to the zero temperature regime
\begin{equation}
\log \mathbb{P}\left(\frac{x_{\mathrm{max}}(T)-x_{\mathrm{edge}}}{w_N}\leq s \right) \simeq - \frac{1}{12} |s|^3
\end{equation}
which is the left tail of the GUE Tracy Widom distribution $F_2(s)$ that indeed
describes the zero temperature fluctuations of the position of the rightmost fermion, 
see Ref. \cite{dean2016noninteracting}.\\

We note that the $N^2$ factor in Eq. \eqref{leftlarge} is reminiscent of the one present in the large deviation of the \textit{pushed} Coulomb gas, see Ref. \cite{majumdar2014top}. Let us recall that the left large deviation rate function for the largest eigenvalue of the GUE ensemble reads
\be
\log P(\lambda_{\rm max},N) \simeq - 2 N^2 \Phi^{RMT}_-\left(\frac{\lambda_{\rm max}-\sqrt{2 N}}{\sqrt{N}}\right)  
\ee
where $\Phi_-^{RMT}(z)$ is calculated in Refs.  \cite{dean2006large, dean2008extreme} and satisfies 
$\Phi_-^{RMT}(z) = \frac{1}{6 \sqrt{2}} \abs{z}^3$ at small $z<0$. One can also check directly that Eq. \eqref{sb2}
for $b \to 0$ also matches the GUE large deviation rate function at small $z$.\\

Finally, note that if we choose to rescale $\tilde{s}$ to take into account the scale of fluctuations $w_N(T)$, we get heuristically
\begin{equation}
\log \mathbb{P}\left(x_{\mathrm{max}}(T)-x_{\mathrm{edge}}\leq \hat{s }\right) \simeq -\frac{2^{13/4}}{15\pi}\frac{N^{3/4}}{T} (-\hat{s})^{5/2}
\end{equation}
The factor $1/T$ now allows us to interpret the large deviation function as a classical Boltzmann factor with a non-trivial, sub-linear $N$ dependence, i.e. as a particle in a $5/2$ power-law potential with sub-extensive amplitude.
%\newpage

%\newpage
\section{Conclusion}
We have presented in this paper two simple methods to compute the large time large deviation tail of the solution of the KPZ equation and have confirmed that for both droplet and Brownian initial conditions, the cumulative distribution has a large deviation form, for $\tilde{s}$ negative and large,
\begin{equation} 
-\log \mathbb{P}(H<t\tilde{s}) \simeq t^2\frac{4}{15\pi}(-\tilde{s})^{5/2}
\end{equation}
where we have shown that these results only depend on the asymptotic density of states of the determinantal process related to the moment generating function of the solution. These methods only use simple tools from determinantal process theory and do not require an involved analysis of the Painlev\'e equation formulation of the related Fredholm determinant, making the generalization straightforward to different kernels that would appear for other initial conditions.\\

We have provided hints of the extension of the cumulant expansion method to arbitrary time and applied it also to a Pfaffian point process - the half-space KPZ problem. In particular, we have observed that the first cumulant alone allows to extract the entire large deviation rate functions at short time that were previously obtained in Refs. \cite{le2016exact, krajenbrink2017exact} by a different method, involving summation of traces of the Fredholm determinant (close in spirit to method A of the present paper). The extension of our method also implies that at any time $t$, as long as $st^{1/3}$ is negative and large enough, we have the following large deviation principle for droplet IC
\vspace{-0.05cm}
\begin{equation}
\label{all_time_large_dev100}
-\log \mathbb{P}(H<st^{1/3}) \simeq \frac{4}{15\pi}t^{1/3}(-s)^{5/2}
\end{equation}
Proceeding to the rescaling $s=t^{-1/3}\hat {s}$, this is indeed the form observed at short time for both droplet and Brownian ICs, see Refs. \cite{le2016exact, MeersonParabola, janas2016dynamical,krajenbrink2017exact,meerson2017randomic}. It is thus likely to conjecture that Eq. \eqref{all_time_large_dev100} also holds for the Brownian IC. For the half-space problem
we have found that the left large deviation rates are consistently equal to half of the full space ones.\\

We also provided exact bounds and a number of conjectures related to the large deviation rate function and the cumulant expansion and presented an interpretation of the large deviations in terms of trapped fermions in quantum mechanics. We hope that these simple methods will provide a further bridge between the different time regimes.
\section*{Acknowledgements}
We acknowledge motivating discussions with Promit Ghosal, Ivan Corwin and Guillaume Barraquand. We are also grateful to Gregory Schehr and Satya N. Majumdar for multiple interactions. This work was partly conducted during the Park City Mathematics Institute 2017 GSS funded by the NSF grant DMS:1441467.
We also acknowledge support from ANR grant ANR-17-CE30-0027-01 RaMaTraF.
%\newpage
\section{Appendix I : calculation of the asymptotics of the deformed Airy kernel and of the Airy kernel at large time}
\label{details_calculation_sine}
In this Appendix, we derive in details the calculation of the asymptotics of the deformed Airy kernel at large time presented in Eq. \eqref{KtAi} and first recall its integral definition, see Refs. \cite{SasamotoStationary,SasamotoStationary2,BCFV}.
\begin{equation}
 \begin{split} 
&  K_{\rm Ai, \Gamma}(v,v') =\\ &\frac{i}{4\pi^2 }\mkern-10mu\int\limits_{-\infty+i \epsilon}^{+\infty+i \epsilon}\int\limits_{-\infty}^{+\infty}\mathrm{d}\eta \mathrm{d}\eta' \frac{\exp\left(i\frac{\eta^3}{3}+iv\eta +i\frac{\eta'^3}{3}+iv'\eta'\right)}{\eta+\eta'}\frac{\Gamma(i t^{-\frac{1}{3}}\eta+w)}{\Gamma(-i t^{-\frac{1}{3}}\eta+w)}  \frac{\Gamma(i t^{-\frac{1}{3}}\eta'+w)}{\Gamma(-i t^{-\frac{1}{3}}\eta'+w)}  
\end{split} 
\end{equation} 
where $\epsilon=0^+$. At large time we proceed to the rescaling $v\to t^{2/3}v$, $\eta\to t^{1/3}\eta$, yielding
\begin{equation}
 \begin{split} 
&  K_{\rm Ai, \Gamma}(vt^{2/3},v't^{2/3}) =\\& \frac{it^{1/3}}{4\pi^2 }\mkern-8mu\int\limits_{-\infty+i \epsilon}^{+\infty+i \epsilon}\int\limits_{-\infty}^{+\infty}\mathrm{d}\eta \mathrm{d}\eta' \frac{\exp\left(it(\frac{\eta^3}{3}+v\eta +\frac{\eta'^3}{3}+v'\eta')\right)}{\eta+\eta'}\frac{\Gamma(i\eta+w)}{\Gamma(-i \eta+w)}  \frac{\Gamma(i \eta'+w)}{\Gamma(-i \eta'+w)}  
\end{split} 
\end{equation} 
As the large parameter $t$ is only present in the exponential, the saddle points are the same as those of the Airy function
\begin{equation}
\eta=\pm \eta_{p} \quad , \quad \eta_p= \sqrt{-v} \quad , \quad \eta'=\pm \eta'_{p} \quad , \quad \eta'_p= \sqrt{-v'} 
\end{equation}
In particular, they force $v$ and $v'$ to be negative so that the kernel does not decrease exponentially. We have four combinations to compute over the pairs of saddle points.
\begin{equation}
 \begin{split} 
&  K_{\rm Ai, \Gamma}(vt^{2/3},v't^{2/3})=\\&  \sum_{\pm_1}\sum_{\pm_2}\frac{it^{-2/3}}{4\pi }\sqrt{\frac{1}{-(\pm_1)(\pm_2)\eta_p \eta'_p}}\frac{\exp\left(\frac{2it}{3}(\mp_1\eta_p^3\mp_2\eta_p'^3)\right)}{\pm_1\eta_p\pm_2\eta_p'}\frac{\Gamma(\pm_1i\eta_p+w)}{\Gamma(\mp_1i \eta_p+w)}  \frac{\Gamma(\pm_2i \eta'_p+w)}{\Gamma(\mp_2i \eta'_p+w)}  
\end{split} 
\end{equation}
We require $\pm_1=(-)\pm_2$ as this choice provides the leading term in time.
\begin{equation}
 \begin{split} 
&  K_{\rm Ai, \Gamma}(vt^{2/3},v't^{2/3})=  \sum_{\pm}\pm \frac{it^{-2/3}}{4\pi } \sqrt{\frac{1}{\eta_p \eta'_p}}\frac{\exp\left(\mp\frac{2it}{3}(\eta_p^3-\eta_p'^3)\right)}{\eta_p-\eta_p'}\frac{\Gamma(\pm i\eta_p+w)}{\Gamma(\mp i \eta_p+w)}  \frac{\Gamma(\mp i \eta'_p+w)}{\Gamma(\pm i \eta'_p+w)}  
\end{split} 
\end{equation}
we finally define $v'=v+\dfrac{\kappa}{t}$, so that $\eta_p'-\eta_p=-\dfrac{\kappa}{t}\dfrac{1}{2\eta_p}$ and observe that  at leading order in $t$ the ratio of $\Gamma$ functions simplifies to 1.
\begin{equation}
 \begin{split} 
  K_{\rm Ai, \Gamma}(vt^{2/3},(v+\frac{\kappa}{t})t^{2/3})&= \sum_{\pm} \pm\frac{it^{1/3}}{2\pi }\frac{\exp\left(\mp i\sqrt{-v}\kappa\right)}{\kappa}\\
  &= \frac{ t^{1/3}}{\pi } \frac{\sin (\sqrt{-v} \kappa)}{\kappa} 
\end{split} 
\end{equation}
which is the formula given in the text. Note that the same formula holds for
the standard Airy kernel $K_{\rm Ai}$ which has the same integral representation without Gamma functions.

\section{Appendix II : Asymptotics of the off-diagonal GOE kernel at large time}
\label{GOE_asymptotic}

Here we show that for $v<0$ 
\be \label{K12asympt} 
K_{12}(vt^{2/3},(v+\frac{\kappa}{t})t^{2/3}) \simeq 
\frac{ t^{1/3}}{\pi } \frac{\sin (\sqrt{-v} \kappa)}{\kappa} 
\ee
We use the definition \eqref{GOE_Airy}. Since we have already shown the asymptotics
\eqref{K12asympt} for $K_{\rm Ai}$ we can now use that
\be
\int_{-\infty}^{y} dr {\rm Ai}(r) \simeq \frac{1}{\sqrt{\pi} |y|^{3/4}} \cos( \frac{2}{3} |y|^{3/2} + \frac{\pi}{4})
\ee 
Hence, from the identity $\int_{-\infty}^{+\infty} dr {\rm Ai}(r)=1$ and the
asymptotics $|{\rm Ai}(x)| \leq_{x \to - \infty} \frac{1}{\sqrt{\pi} |x|^{1/4}}$
one sees that the additional terms in \eqref{GOE_Airy} are negligible since
\be
\lim_{t \to +\infty} t^{-1/3} {\rm Ai}( v t^{2/3}) \int_{-\infty}^{0} dr {\rm Ai}(r
+ (v+\frac{\kappa}{t})t^{2/3}))  = 0
\ee 
\section{Appendix III : Calculation of the long time estimate of the Fredholm determinant}
\label{appendix2}
We present in this Appendix the details of the calculation of the long estimate of the Fredholm determinant presented in Section \ref{method1}. Using the identity $\log \mathrm{Det}=\mathrm{Tr}\log$, and expanding the logarithm into a series, the Fredholm determinant can be computed as 
\bea\label{Q_start_supp}
\log Q_t(s) = - \sum_{p=1}^\infty \frac{1}{p} {\rm Tr}\, \bar{K}^p_{t,s} 
\eea
where we recall the definition of the trace of a Kernel
\begin{equation}
{\rm Tr} \; {\bar K}_{t,s}^p = \int_{-\infty}^\infty dv_1 \int_{-\infty}^\infty dv_2 \ldots \int_{-\infty}^\infty dv_p K_{\rm Ai}(v_1,v_2) .. K_{\rm Ai}(v_p,v_1)  \sigma_{t,s}(v_1) \ldots  \sigma_{t,s}(v_p)
\end{equation}
Proceeding to the proposed rescaling of Section \ref{choice_scaling}, we obtain
 \be\label{trace_Kp2}
 \begin{split}
{\rm Tr} \; {\bar K}_{t,s}^p =&\,  t^{2p/3} \int_{-\infty}^\infty dv_1 \int_{-\infty}^\infty dv_2 \ldots \int_{-\infty}^\infty dv_p \, K_{\rm Ai}\left(v_1 t^{2/3} ,v_2 t^{2/3} \right) \ldots K_{\rm Ai}\left(v_p t^{2/3} ,v_1 t^{2/3} \right) \\
& \sigma(t(v_1-\tilde s)) \dots \sigma(t(v_p-\tilde s))
\end{split}
\ee
One can then use the asymptotic expansion of the Airy kernel in the large time limit, see Eq. \eqref{KtAi} and Section \ref{details_calculation_sine}, 
\bea
 K_{\rm Ai}\left(v_1 t^{2/3} ,(v_1 + \frac{\kappa}{t})t^{2/3}\right)  \underset{t\gg 1}\simeq 
\frac{ t^{1/3}}{\pi } \frac{\sin (\sqrt{|v_1|} \kappa)}{\kappa} \Theta(-v_1) \,.
\eea 
On the other hand, for $v_1 > 0$, the Airy kernel vanishes exponentially in the long time limit
and therefore only the region where all the $v_i$ are negative needs to be considered
in Eq. \eqref{trace_Kp2}. Hence for $p \geq 2$, 
separating the center of mass coordinate (which we take as $v_1$) 
and the $p-1$ relative coordinates $v_j=v_{j-1}+ \frac{\kappa_j}{t}$ 
we obtain after rescaling all $\kappa$'s by $\sqrt{v_1}$
\be
\begin{split}
&{\rm Tr} \bar{K}^p 
\simeq \frac{t}{\pi^p}   \int_{\tilde{s}}^0 dv_1 \sqrt{|v_1|} \int_{t\tilde{s}\sqrt{|v_1|}}^{-t\tilde{s}\sqrt{|v_1|}} d\kappa_1  \ldots \int_{t\tilde{s}\sqrt{|v_1|}}^{-t\tilde{s}\sqrt{|v_1|}} d\kappa_p\\
& \prod_{i=1}^p \sigma(t(v_1-\tilde s)+\sum_{j=2}^{i}\frac{\kappa_j}{\sqrt{|v_1|}})   \frac{\sin ( \kappa_1)}{\kappa_1} \frac{\sin (  \kappa_2)}{\kappa_2} \ldots \frac{\sin ( \kappa_p)}{\kappa_p}
\delta(\kappa_1+\kappa_2+\dots+\kappa_p) 
\end{split}
\ee
We have reduced the range of integration of $v_1$ due to the fact that the weight function vanishes exponentially for $v_1<\tilde{s}$. If $\tilde{s}$ is positive, then this unfortunately competes with the exponential decrease of the Airy kernel. For negative $\tilde{s}$, there exists a small interval $[\tilde{s},0]$ where things do not vanish exponentially. As we are interested in the left tail of the distribution, we are indeed in the case $\tilde{s}<0$. Note that from this expression, the symmetry $\lbrace \kappa \rbrace \leftrightarrow \lbrace -\kappa \rbrace$ is preserved.\\

We now want to evaluate the leading order of the Fredholm determinant. We make the \textit{hypothesis} that we can neglect the term $\sum_{j=2}^{i}\frac{\kappa_j}{\sqrt{|v_1|}}$ in the Fermi factors, then the traces can be rewritten as 
\begin{equation}
\begin{split}
{\rm Tr} \bar{K}^p 
\simeq& \frac{t}{\pi^p}   \int_{\tilde{s}}^0 dv_1 \sqrt{|v_1|}  \sigma(t(v_1-\tilde s))^p  \int_{t\tilde{s}\sqrt{|v_1|}}^{-t\tilde{s}\sqrt{|v_1|}} d\kappa_1  \ldots \int_{t\tilde{s}\sqrt{|v_1|}}^{-t\tilde{s}\sqrt{|v_1|}} d\kappa_p \\ &\frac{\sin ( \kappa_1)}{\kappa_1} \frac{\sin (  \kappa_2)}{\kappa_2} \ldots \frac{\sin ( \kappa_p)}{\kappa_p}
\delta(\kappa_1+\kappa_2+\dots+\kappa_p) 
\end{split}
\end{equation}
In order to deal with the coupling between the $\kappa$'s, we express the $\delta$ functions in Fourier space.
\begin{equation}
\begin{split}
I_p&= \int_{t\tilde{s}}^{-t\tilde{s}} d\kappa_1  \ldots \int_{t\tilde{s}\sqrt{|v_1|}}^{-t\tilde{s}\sqrt{|v_1|}} d\kappa_p \frac{\sin ( \kappa_1)}{\kappa_1} \frac{\sin (  \kappa_2)}{\kappa_2} \ldots \frac{\sin ( \kappa_p)}{\kappa_p}
\delta(\kappa_1+\kappa_2+\dots+\kappa_p) \\
&= \int_{-\infty}^\infty \frac{dk}{2 \pi}\int_{t\tilde{s}\sqrt{|v_1|}}^{-t\tilde{s}\sqrt{|v_1|}} d\kappa_1  \ldots \int_{t\tilde{s}\sqrt{|v_1|}}^{-t\tilde{s}\sqrt{|v_1|}} d\kappa_p  \frac{\sin ( \kappa_1)}{\kappa_1} \frac{\sin (  \kappa_2)}{\kappa_2} \ldots \frac{\sin ( \kappa_p)}{\kappa_p}  e^{ik\sum_{j=1}^p \kappa_j} \\
&= \int_{-\infty}^\infty \frac{dk}{2 \pi} \left( \mathrm{Si}(-t\tilde{s}\sqrt{|v_1|}(1+k))+\mathrm{Si}(-t\tilde{s}\sqrt{|v_1|}(1-k)) \right)^p\\
\end{split}
\end{equation}
where $\mathrm{Si}(x)=\int_{0}^x\mathrm{d}s \, \frac{\sin(s)}{s}$. Using the limit for $\sqrt{|v_1|}$ fixed
\begin{equation}
\lim_{-t\tilde{s}\to +\infty}\mathrm{Si}(-t\tilde{s}\sqrt{|v_1|}(1+k))=\dfrac{\pi}{2}\, \mathrm{sgn}(1+k),
\end{equation}
we deduce for any $p\geq 1$
\begin{equation}
\lim_{-t\tilde{s}\rightarrow +\infty}\left( \mathrm{Si}(-t\tilde{s}\sqrt{|v_1|}(1+k))+\mathrm{Si}(-t\tilde{s}\sqrt{|v_1|}(1-k)) \right)^p=\pi^p\; \mathds{1}_{k\in [-1,1]}
\end{equation}
where $\mathds{1}_{k\in [-1,1]}$ is the indicator function of $[-1,1]$, allowing us to obtain $I_p=\pi^{p-1}$ as in Refs. \cite{le2016exact, krajenbrink2017exact}. The trace is therefore simplified onto
\bea{\rm Tr} \bar{K}^p 
\simeq \frac{t}{\pi}   \int_{\tilde{s}}^0 dv_1 \sqrt{|v_1|}  \sigma(t(v_1-\tilde s))^p 
\eea
Doing the summation over $p$, we obtain as in Section \ref{method1}
\begin{equation}
\begin{split}
\log Q_t(s)&=-\frac{t}{\pi} \int_{\tilde{s}}^0 dv_1 \sqrt{|v_1|}  \sum_{p=1}^\infty \frac{ \sigma(t(v_1-\tilde s))^p}{p}\\
&=-\frac{t}{\pi} \int_{\tilde{s}}^0 dv_1 \sqrt{|v_1|} \ln(1+e^{t(v_1-\tilde{s})})\\
&\simeq -\frac{t^2§}{\pi} \int_{\tilde{s}}^0 dv_1 \sqrt{|v_1|} (v_1-\tilde{s})\\
&\simeq -\frac{4t^2§}{15\pi}(-\tilde{s})^{5/2}
\end{split}
\end{equation} 
%\newpage
\section{Appendix IV : Fredholm determinant, cumulant expansion and BCH formula}
\label{appendix3}
\subsection{Cumulant expansion}
For a set of points $\lbrace a_i \rbrace_{i \in \mathbb{N}}$ following a determinantal point process with kernel $K$, we are interested in evaluating the general quantity $
\mathbb{E}_K \left[\exp\left(-\sum_i \varphi(a_i)\right)\right]$
for any function $\varphi$. As derived in the text and recalling results from Refs. \cite{borodin2016moments, johansson2005random}, we have the identity
  \begin{equation} 
 \begin{split}
\log \mathbb{E}_K \left[ \exp\left(-\sum_{i=1}^{\infty} \varphi(a_i)\right)\right]&= -\sum_{p=1}^{+\infty} \frac{1}{p}\mathrm{Tr}[(1-e^{-\varphi})K]^p\\
\end{split}
 \end{equation}
Expanding $1-e^{-\varphi}$ to the third order allows us to obtain the first three cumulants.
 \begin{enumerate}
 \item \textit{p=1 }
 \begin{equation}
-\mathrm{Tr}[(1-e^{-\varphi})K]= -\mathrm{Tr}(\varphi K)+\frac{1}{2}\mathrm{Tr}(\varphi^2 K)-\frac{1}{6}\mathrm{Tr}(\varphi^3 K)
\end{equation}  
 \item \textit{p=2 }
 \begin{equation}
 -\frac{1}{2}\mathrm{Tr}[(1-e^{-\varphi})K]^2=-\frac{1}{2}\mathrm{Tr}(\varphi K \varphi K)+\frac{1}{2}\mathrm{Tr}(\varphi K \varphi^2 K)
 \end{equation}
 \item \textit{p=3 }
 \begin{equation}
 -\frac{1}{3}\mathrm{Tr}[(1-e^{-\varphi})K]^3=-\frac{1}{3}\mathrm{Tr}(\varphi K \varphi K \varphi	K)
 \end{equation}
 \end{enumerate}
The $n$-th cumulant $\kappa_n(\varphi)$ is defined as $n!$ times the term of order $\varphi^n$ in this expansion
\begin{equation} \label{defcum} 
 \log \mathbb{E}_K \left[ \exp\left(-\sum_{i=1}^\infty \varphi(a_i)\right)\right]=\sum_{n=1}^\infty \frac{\kappa_n(\varphi)}{n!}
\end{equation}
Grouping the various terms, we end up having  
 \begin{equation}
  \label{third_order_expansion}
 \begin{split}
& \kappa_1(\varphi)=-\mathrm{Tr}(\varphi K), \qquad \kappa_2(\varphi)=\mathrm{Tr}(\varphi^2 K)-\mathrm{Tr}(\varphi K \varphi K), \\
& \kappa_3(\varphi)=-\mathrm{Tr}(\varphi^3 K) +3 \mathrm{Tr}(\varphi K \varphi^2 K)-2 \mathrm{Tr}(\varphi K \varphi K \varphi	K)
 \end{split}
 \end{equation}
The formula for the general cumulant can be found in Ref. \cite{johansson2015gaussian} where similar problems of linear statistics have been studied
\begin{equation}
\kappa_n(\varphi)=\sum_{l=1}^n \frac{(-1)^{n+l+1}}{l}\sum_{\substack{m_1,\dots,m_l \geq 1 \\ m_1+\dots+m_l=n}}\frac{n!}{m_1!\dots m_l !}\mathrm{Tr}(\varphi^{m_1}K\varphi^{m_2}K\dots \varphi^{m_l}K)
\end{equation}

 \subsection{Parallel with the BCH formula}
Noticing that $\mathrm{Tr}(K)=\mathrm{Tr}(K^2)$ by the reproducing property, we can rewrite the second cumulant in another way, using the cyclicity of the trace
 \begin{equation}
 \begin{split}
\mathrm{Tr}(\varphi^2 K)-\mathrm{Tr}(\varphi K \varphi K)&=\mathrm{Tr}(\varphi KK\varphi )-\mathrm{Tr}(\varphi K \varphi K)\\
&= \mathrm{Tr}(\varphi K[K\varphi-\varphi K] )
\end{split}
 \end{equation}
 It is possible to define a commutator $[K,\varphi]=K\varphi-\varphi K $ so that  $ [K,\varphi] (a,b)= K(a,b)\left(\varphi(b)-\varphi(a)\right)$ and to call some extension of the Baker-Campbell-Hausdorff (BCH) formula for determinantal processes. In the same way, the third cumulant gives
  \begin{equation}
\mathrm{Tr}(\varphi^3 K) -3 \mathrm{Tr}(\varphi K \varphi^2 K)+2 \mathrm{Tr}(\varphi K \varphi K \varphi	K)= \mathrm{Tr}\left(\varphi K\left(2[\varphi,K]^2+[\varphi,[\varphi,K]]\right)\right)
 \end{equation}
 Keeping the first three orders, we have
 \begin{equation}
 \begin{split}
  \log\mathbb{E}_K \left[\exp\left(-\sum_{i=1}^\infty \varphi(a_i)\right)\right]&=-\mathrm{Tr}\left(\varphi K\, \left(1+\frac{1}{2}[\varphi,K]+\frac{1}{3}[\varphi,K]^2+\frac{1}{6}[\varphi,[\varphi,K]]\, \right) \right)\\
  &+\text{higher order cumulants}
  \end{split}
 \end{equation}
At the moment the link with the BCH formula is purely conjectural. A direction would be to investigate whether an analogue structure holds for higher order cumulants. We finally recall for completeness the BCH formula, see Ref. \cite{suzuki1977convergence}, for any square matrices $X$ and $Y$
 \begin{equation}
 \log\left(\exp X\exp Y\right) = X + Y + \frac{1}{2}[X,Y] +
\frac{1}{12}\left ([X,[X,Y]] +[Y,[Y,X]]\right )+\dots
 \end{equation}
%\newpage

\section{Appendix V : Conditional determinantal point process} \label{app:conditional}
In this appendix we show, in a very simple way, that a determinantal point process
conditioned to its largest point is again a determinantal point process with a new
kernel. More elaborate proofs can be found in \cite{BufetovDeterminantal}. 

Let us choose a set $\lbrace a_i \rbrace $ following a determinantal point process with kernel $K$. We condition the process on the largest element being smaller than $s$, i.e. $a_1<s$. We find below that this
new point process is again a determinantal point process, 
with a new kernel involving the resolvant of $K$, and equal to $(I-P_s)K(I-P_s K)^{-1}$,
where we have defined the projector $P_s(a)=\theta(a>s)$.
Calling $\rho(a,s)$ the associated conditional density , we have 
\begin{equation}
\label{conditional_density}
\rho(a,s)=\left[(I-P_s)K(I-P_s K)^{-1}\right]_{aa}
\end{equation}
\begin{proof}
For any point process and any function $f$, the density defined as $\rho(a,s)=\mathbb{E}\left[ \sum_{i=1}^{+\infty} \delta(a-a_i)\mid a_1<s \right]$ is obtained as
\begin{equation}
\label{conditional_process}
\begin{split}
\mathbb{E}\left[\prod_{i=1}^{+\infty} (1+\varepsilon f(a_i))\mid a_1<s \right]&=1+\varepsilon \, \mathbb{E} \left[ \sum_{i=1}^{+\infty} f(a_i)\mid a_1<s\right]+\mathcal{O}(\varepsilon^2)\\
&= 1+\varepsilon \int_\Omega \mathrm{d}a \rho(a,s) f(a)+\mathcal{O}(\varepsilon^2)
\end{split}
\end{equation}
For the conditioned process, using the Fredholm determinant representation of the expectation value \eqref{conditional_process} and Bayes formula, we obtain 
\begin{equation}
\begin{split}
\mathbb{E}\left[\prod_{i=1}^{+\infty} (1+\varepsilon f(a_i)) \mid a_1<s \right]&=\frac{\mathbb{E}\left[\prod_{i=1}^{+\infty} (1+\varepsilon f(a_i)) \theta(a_i<s)\right]}{\mathbb{P}(a<s)}\\
&=\frac{\mathrm{Det}\left[I+[(1+\varepsilon f)\theta(a<s)-I]K \right]}{\mathrm{Det}\left[ I-\theta(a>s)K \right]}\\
&=\frac{\mathrm{Det}\left[I-\theta(a>s)K+\varepsilon f \theta(a<s)K \right]}{\mathrm{Det}\left[ I-\theta(a>s)K \right]}\\
& = \mathrm{Det}\left[I + \varepsilon f (I-P_s)K(I-P_s K)^{-1}\right] \\
&= 1+\varepsilon \, \mathrm{Tr}(f\theta(a<s)K[I-\theta(a>s)K]^{-1})+\mathcal{O}(\varepsilon^2)
\end{split}
\end{equation}
Identifying the first order in $\varepsilon$ for any function $f$ yields the formula \eqref{conditional_density},
and the line before last shows that it is a determinantal process with the kernel defined above.
\end{proof}

%\newpage


\begin{thebibliography}{99}
\bibitem{footnote100}
This is equivalent to use everywhere the following units of space, time and heights\\
$ x^*=(2 \nu)^3/(D \lambda_0^2), \quad t^*=2(2 \nu)^5/(D^2 \lambda_0^4) , \quad 
h^*=\frac{2 \nu}{\lambda_0}$
\end{thebibliography}
\end{document}